\hfuzz 2pt
\font\titlefont=cmbx10 scaled\magstep1
\magnification=\magstep1

\null
\vskip .5cm
\centerline{\titlefont IRREVERSIBILITY AND DISSIPATION}
\smallskip
\centerline{\titlefont IN NEUTRAL B-MESON DECAYS}
\vskip 1.8cm
\centerline{\bf F. Benatti}
\smallskip
\centerline{Dipartimento di Fisica Teorica, Universit\`a di Trieste}
\centerline{Strada Costiera 11, 34014 Trieste, Italy}
\centerline{and}
\centerline{Istituto Nazionale di Fisica Nucleare, Sezione di 
Trieste}
\vskip 1cm
\centerline{\bf R. Floreanini}
\smallskip
\centerline{Istituto Nazionale di Fisica Nucleare, Sezione di 
Trieste}
\centerline{Dipartimento di Fisica Teorica, Universit\`a di Trieste}
\centerline{Strada Costiera 11, 34014 Trieste, Italy}
\vskip 1cm
\centerline{\bf R. Romano}
\smallskip
\centerline{Dipartimento di Fisica Teorica, Universit\`a di Trieste}
\centerline{Strada Costiera 11, 34014 Trieste, Italy}
\centerline{and}
\centerline{Istituto Nazionale di Fisica Nucleare, Sezione di 
Trieste}
\vskip 2.5cm
\centerline{\bf Abstract}
\smallskip
\midinsert
\narrower\narrower\noindent
The propagation and decay of neutral $B$-mesons can be described in terms
of quantum dynamical semigroups; they provide generalized time-evolutions that
take into account possible non-standard effects leading to loss of phase coherence
and dissipation. These effects can be fully parametrized in terms of
six phenomenological constants. A detailed analysis of selected 
$B$-meson decays shows that present and future dedicated
experiments, both at colliders and $B$-factories, will be able to put
stringent bounds on these non-standard parameters. 
\endinsert
\bigskip
\vfill\eject

{\bf 1. INTRODUCTION}
\medskip

The study of physics related to $b$-flavored hadrons appears to be 
a very promising testing ground for various fundamental aspects of the
standard model. Indeed, a considerable effort has recently been devoted
to the realization of specific experimental programs in order to probe with high
accuracy the $b$-sector of the theory. The so-called $B$-factories are
already collecting data, while experiments at colliders will start running
in the near future and other will be constructed in the coming years, so that
very precise data, in particular on $CP$-violating phenomena, will be
available for comparison with the theoretical predictions.[1-3]

The neutral $B$-meson system is in this respect a unique laboratory:
the $B^0$-$\overline{B^0}$ mixing and the induced interference phenomena
allow, at least in principle, an accurate analysis of very small effects.
Besides $CP$-violating phenomena, these could also include effects induced
by physics beyond the standard model, like those predicted by gran-unified
theories, supersymmetry and even fundamental dynamics of strings and branes.

In the following, we shall devote our attention to the study of the
non-standard effects that are induced in neutral $B$-meson physics
by a suitable generalization of the familiar,
effective time-evolution of ordinary quantum mechanics.
This generalized dynamics takes into account possible phenomena leading
to irreversibility and dissipation that could affect various
$B$-meson observables; we shall discuss in detail to what extent
present and future dedicated experiments will be able to detect such effects.

Our approach will be phenomenological in nature; in particular, we shall not
need to discuss in detail how the generalized time-evolution originates
from the fundamental dynamics. In fact, the effective description we shall 
discuss turns out to be largely independent from the microscopic phenomena 
responsible for the appearance of the new effects.

This phenomenological point of view is perfectly consistent:
not all dynamics that generalize the familiar, unitary time-evolution
of quantum mechanics are in fact physically acceptable; basic
requirements need to be enforced.
The new time-evolution should always transform, in all physical situations, 
$B$-meson states into $B$-meson states, allow forward in time composition law,
while increasing the system entropy. As we shall see, these conditions uniquely
fix the form of the generalized dynamics, that turns out to be parametrized
in terms of six, new phenomenological constants. In more mathematically
precise terms, the generalized time-evolution takes 
the form of a quantum dynamical semigroup.[4-6]

A physical instance in which evolutions of this type are encountered is given
by the study of open quantum systems.[4-8] 
Quite in general, an open system can be
considered as a subsystem $S$ in interaction with a large environment $E$.
Although the complete system $S+E$ evolves in time with the standard
unitary operator of quantum mechanics, the subdynamics of $S$ alone,
obtained by a suitable integration over the environment degrees of
freedom, is rather involved, showing in general irreversibility and memory
effects. However, when the interaction between the subsystem and the
environment can be considered to be weak, the dynamics of $S$ simplifies;
it can be represented by linear maps, local in time, that 
take precisely the form of a quantum dynamical semigroup.

This description is very general and can be applied to model different
physical situations. It has been originally developed in the framework of
quantum optics,[8-10] but it has also been successfully used to study statistical
models,[4-6] the interaction of a microsystem with a measuring apparatus,[11-13]
non-standard effects in different interferometric experiments involving
elementary particles (neutrons,[14] neutrinos [15] and photons [16]).%
\footnote{$^\dagger$}{Propagation of neutrinos in
thermal environments can also be described by equations of semigroup
form; for details, see [17, 18] and the discussion in [15]}
Furthermore, it provides a general framework for
the analysis of dissipative effects in the evolution and decay of
neutral meson systems.[19, 20, 21-25] (For early investigations, see [26].)

The original physical motivation for such investigations 
came from quantum gravity: due to the quantum fluctuation of
the gravitational field and the presence of virtual black holes, space-time
should lose its continuum character at Planck's scale, and this could lead
to loss of quantum coherence.[26-33] From a more fundamental point of view,
also the dynamics of strings and branes could act as an effective environment,
inducing non-standard, dissipative effects at low energies.[34, 35]
Fortunately, as already mentioned, the details of the microscopic, ``stringy''
dynamics are not needed for the effective description
of the dissipative phenomena in terms of quantum dynamical semigroups.[35]
In this respect, this phenomenological approach provides a universal
framework for the study of quantum decoherence effects.

In the case of neutral kaon system, the analysis of the
dissipative phenomena has been pursued
in detail, and upper bounds on some of the constants parametrizing the 
new effects have been obtained using available experimental data.[21, 22]
Improvements of these results are expected from the experimental study
of correlated $K$-mesons at $\phi$-factories.\hbox{[23, 24]}

For the case of the neutral $B$-meson system, a preliminary analysis
of the relevance of dissipative effects on semileptonic decays has been
presented in [25]. There, a particular phase choice for the $B$ states
has been adopted; it allowed treating indirect $CP$-violating effects
on the same footing as the decoherence phenomena.

In the following, a much more general and complete discussion will be
presented. Using a manifestly phase-invariant formalism
that makes no commitment on the magnitude 
of $CP$ (and $CPT$) violating effects,
the impact of the irreversible, dissipative phenomena on neutral 
$B$-meson decays will be analyzed and discussed in detail,
using both semileptonic and selected hadronic decay channels.
We shall first study decays of single mesons, relevant for
experiments at colliders, and then discuss the case of correlated
mesons at $B$-factories. In both cases, the new, dissipative phenomena
modify in a specific and characteristic way the various $B$-meson 
observables, so that the presence of dissipation
can be probed quite independently from other non-standard effects.%
\footnote{$^\ddagger$}{For recent studies on possible violations of the
$CPT$ symmetry and other, unconventional phenomena in $B$-decays, 
see [36-41] and references therein.}
Although more specific studies, that include precise analysis of acceptance
and efficiency of the various detectors, are certainly necessary,
our investigation clearly indicates that present and
future $B$-meson experiments should be able to put stringent
limits on the parameters that describe non-standard, dissipative phenomena.

\vfill\eject

{\bf 2. QUANTUM DYNAMICAL SEMIGROUPS}
\medskip

The familiar effective description of the propagation and decay 
of neutral $B$-mesons requires the introduction of a two-dimensional 
Hilbert space;%
\footnote{$^\dagger$}{We shall limit our considerations to $B_d$-mesons,
although most of the discussions below could be applied to $B_s$-mesons
as well.}
the time-evolution is then realized by a linear 
transformation on the elements of this space.[1, 2] In presence of
dissipation however, a more general formalism is needed,
in which $B$-meson states are represented by density matrices.
These are hermitian, positive operators
({\it i.e.} with positive eigenvalues), with constant trace
(at least for unitary evolutions). In the 
$|B^0\rangle$, $|\overline{B^0}\rangle$ basis, any $B$-meson
state can then be written as
$$
\rho=\left[\matrix{
\rho_1&\rho_3\cr
\rho_4&\rho_2}\right]\ , \eqno(2.1)
$$
where $\rho_4\equiv\rho_3^*$, and $*$ signifies complex conjugation.

As explained in the introductory remarks, our analysis is based on the
hypothesis that the time evolution of the state $\rho$ is
realized in terms of linear transformations: they are generated
by an equation that extends the familiar quantum mechanical one,
being of the form:
$$
{\partial\rho(t)\over\partial t}=-iH\, \rho(t)+i\rho(t)\, H^\dagger 
+L[\rho] .\eqno(2.2)
$$
The first two pieces on the r.h.s. give the standard hamiltonian
contribution, while $L$ is a linear map that encodes possible dissipative,
non-standard effects.

The effective hamiltonian $H$ includes a non-hermitian part,
$$
H= M-{i\over 2}{\mit\Gamma}\ ,\eqno(2.3)
$$
with $M$ and $\mit\Gamma$ hermitian matrices,
that characterizes the natural width of the states.
The entries of $H$ can be expressed in terms of its eigenvalues:
$\lambda_S=m_S-{i\over 2}\gamma_S$, $\lambda_L=m_L-{i\over 2}\gamma_L$,
and the complex parameters $p_S$, $q_S$, $p_L$, $q_L$, appearing in the
corresponding (right) eigenstates, 
$$
\eqalign{
&|B_S\rangle=p_S\, |B^0\rangle + q_S\, |\overline{B^0}\rangle\ ,\qquad\quad
|p_S|^2 + |q_S|^2=1\ ,\cr
&|B_L\rangle=p_L\, |B^0\rangle - q_L\, |\overline{B^0}\rangle\ ,\qquad\quad
|p_L|^2 + |q_L|^2=1\ .}
\eqno(2.4)
$$
We use here a notation that follows the conventions usually adopted for the
neutral kaon system. The two states in (2.4) are expected to have a 
negligible width difference,
$$
\Delta\Gamma\ll \Gamma\ ,\qquad \Delta\Gamma=\gamma_S-\gamma_L\ ,
\qquad \Gamma={\gamma_S+\gamma_L\over2}\ ,\eqno(2.5)
$$
so that they may be more conveniently distinguished via their mass difference,
$\Delta m=m_L - m_S$,
rather than their different lifetimes. 
For this reason, the notation $|B_L\rangle$, for the
light meson, and $|B_H\rangle$, for the heavy partner are therefore 
sometimes used instead of those in (2.4).[1, 2]

The effective hamiltonian $H$ can be diagonalized
using the similarity transformation induced by (2.4):
$$
H=V\, H_0\, V^{-1}\ ,
\eqno(2.6)
$$
with
$$
V=\left[\matrix{p_S&\phantom{-}p_L\cr
                q_S&-q_L}\right]\ ,\ \qquad
H_0=\left[\matrix{\lambda_S&0\cr
                0&\lambda_L}\right]\ .
\eqno(2.7)
$$
Then, one can write:
$$
H\equiv\left[\matrix{H_1&H_3\cr
                     H_4&H_2}\right]=
{1\over r_S+r_L}\left[\matrix{r_S\lambda_S+r_L\lambda_L&
r_S r_L(\lambda_S-\lambda_L)\cr
\lambda_S-\lambda_L&r_L\lambda_S+r_S\lambda_L}\right]\ ,
\eqno(2.8)
$$
where we find useful to introduce the two ratios:
$$
r_S={p_S\over q_S}\ ,\ \qquad r_L={p_L\over q_L}\ .
\eqno(2.9)
$$
They can be conveniently expressed as
$$
r_S=\sqrt\sigma\, \sqrt{1+\theta\over1-\theta}\ ,\ \qquad
r_L=\sqrt\sigma\, \sqrt{1-\theta\over1+\theta}\ ,\ \qquad
\eqno(2.10)
$$
in terms of the two complex parameters
$$
\sigma=r_S r_L\ ,\ \qquad \theta={r_S-r_L\over r_S+r_L}\ ,
\eqno(2.11)
$$
that signal $T$ and $CPT$ violating effects. Indeed, from (2.8) one has:
$$
\sigma={H_3\over H_4}\ ,\ \qquad \theta={H_1-H_2\over\lambda_S-\lambda_L}\ ,
\eqno(2.12)
$$
so that $T$ (and $CP$) invariance is broken when $|\sigma|\neq 1$,
while $CPT$ (and $CP$) invariance is lost for $\theta\neq0$.[1, 2]

Even in absence of the additional piece $L[\rho]$ in (2.2), probability
is not conserved during the time evolution: $d {\rm Tr}[\rho(t)]/dt\leq0$.
This is due to the presence of a non-hermitian part in the effective 
hamiltonian $H$. On the other hand, loss of phase 
coherence shows up only when the piece $L[\rho]$ is nonvanishing:
it produces dissipation and possible transitions from pure states to
mixed states.

As already mentioned before, not all maps $L[\rho]$ gives rise
to integrated time evolutions, $\gamma_t: \rho(0)\mapsto\rho(t)$,
that are physically acceptable. Quite in general, 
the one parameter (=time) family of linear maps
$\gamma_t$ should transform $B$-meson states into
$B$-meson states, and therefore should map an initial density matrix
into a density matrix; further, it should
have the property of increasing the (von Neumann)
entropy, $S=-{\rm Tr}[\rho(t)\,\ln\rho(t)]$, of obeying the semigroup
composition law, $\gamma_t[\rho(t')]=\rho(t+t')$, for $t,\ t'\geq0$,
of preserving the positivity of $\rho(t)$ for all times.
Actually, for the physical consistency of the formalism
in the case of correlated systems, one has to demand that
the time evolution $\gamma_t$ be completely positive.[4-6]
Once these properties are taken into account, the
form of $L[\rho]$ results uniquely fixed and the family
of maps $\gamma_t$ represents a so-called quantum dynamical semigroup.

The linear map $L[\rho]$ can be fully parametrized in terms of
six real phenomenological constants,
$a$, $b$, $c$, $\alpha$, $\beta$ and $\gamma$,
with $a$, $\alpha$, $\gamma$ non negative,
satisfying the following inequalities:[21, 23]
$$
\eqalign{
&2\,R\equiv\alpha+\gamma-a\geq0\ ,\cr
&2\,S\equiv a+\gamma-\alpha\geq0\ ,\cr
&2\,T\equiv a+\alpha-\gamma\geq0\ ,\cr
&RST-2\, bc\beta-R\beta^2-S c^2-T b^2\geq 0\ ,
}\hskip -1cm
\eqalign{
&RS-b^2\geq 0\ ,\cr
&RT-c^2\geq 0\ ,\cr
&ST-\beta^2\geq 0\ ,\cr
&\phantom{\beta^2}\cr
}\eqno(2.13)
$$
direct consequence of the property of complete positivity.
A convenient explicit expression for $L[\rho]$ can be obtained
by introducing a vector notation for the matrix $\rho$,
rewritten as a four-dimensional column vector $|\rho\rangle$,
with components $(\rho_1,\rho_2,\rho_3,\rho_4)$.
In this way, the map $L[\rho]$ can be represented by
a $4\times 4$ hermitian matrix $\cal L$, acting on $|\rho\rangle$:
$$
{\cal L}=\left[\matrix{-a&\phantom{-}a&-(c-ib)&-(c+ib)\cr
                       \phantom{-}a&-a&\phantom{-}c-ib&\phantom{-}c+ib\cr
                       -(c+ib)&c+ib&-(\alpha+\gamma)&\alpha-\gamma-2i\beta\cr
                       -(c-ib)&c-ib&\alpha-\gamma+2i\beta&-(\alpha+\gamma)\cr}
					   \right]\ .\eqno(2.14)
$$

Among the physical requirements that the complete time evolution
$\gamma_t$ should satisfy, complete positivity
is perhaps the less intuitive. It is often dismissed, in favor of the more obvious
simple positivity. 
Simple positivity is in fact generally enough to guarantee that
the eigenvalues of the density matrix $\rho(t)$ 
remain positive at any time; this requirement is obviously crucial
for the consistency of the formalism, in view of the interpretation of the
eigenvalues of $\rho(t)$ as probabilities.

Complete positivity is a stronger property, in the sense that
it assures the positivity of the density matrix describing the
states of a larger system, involving the coupling
with an extra, auxiliary finite-dimensional system. 
Although trivially satisfied by standard
quantum mechanical (unitary) time-evolutions, 
the requirement of complete positivity
seems at first a mere technical complication.
Nevertheless, it turns out to be essential in properly
treating correlated systems, like the two $B$-meson coming
from the $\Upsilon(4S)$ resonance;[42] it assures the absence of 
unphysical effects, like the appearance of negative probabilities,
that could occur for just simply positive dynamics
(see the discussion in Sect.7).

A further comment on the expression (2.14) of the dissipative part of the 
evolution equation (2.2) is in order. The entries of the matrix (2.14)
have been written in a specific choice of basis in the two-dimensional
Hilbert space, the one for which the state $\rho$ takes the form (2.1).
If one performs a (unitary) transformation on the basis vectors
$|B^0\rangle$, $|\overline{B^0}\rangle$, also the six parameters appearing
in (2.14) will in general change to new ones 
$a'$, $b'$, $c'$, $\alpha'$, $\beta'$, and $\gamma'$,
expressed as linear combinations of the old ones.
Despite this, one can check that the inequalities in (2.13) are form-invariant;
in other terms, the transformed dissipative parameters still obey (2.13),
provided the old ones do.%
\footnote{$^\dagger$}{For a generic, non-unitary change of basis, as the one
discussed in Appendix A, also the form of the conditions (2.13) in general
changes, while always assuring the fulfillment of
the property of complete positivity.}

This discussion clearly illustrates that, although the equation (2.2)
has been written in a fixed, specific $B$-meson basis, the notion
of complete positivity is basis-independent, as is the description of
the dissipative effects: if the contribution $\cal L$ in (2.14) is non vanishing
in one basis, it is non vanishing in any basis. This result has to be
expected from the general theory of quantum dynamical semigroups,
that can be formulated in a basis-independent way,[4-6] and will be 
made apparent in the discussion of the coming sections.

\vskip 1cm

{\bf 3. THE EFFECTIVE TIME-EVOLUTION}
\medskip

The behaviour in time of physical observables related to the various
$B$-meson decay channels can be obtained by solving the evolution equation
(2.2) for an arbitrary initial state $\rho(0)$. This results in the study
of a system of linear differential equations for the entries of
the density matrix $\rho$ in (2.1). 

It is convenient to use the vector notation introduced in the
previous section, and write the matrix $\rho$ as the four-dimensional
vector $|\rho\rangle$. Then, the evolution equation (2.2)
takes the form of a Schr\"odinger (or diffusion) equation:
$$
{d\over d t}|\rho(t)\rangle={\cal K}\, |\rho(t)\rangle\equiv
\big[{\cal H}+{\cal L}\big]\, |\rho(t)\rangle\ ,
\eqno(3.1)
$$
where $\cal H$ is the $4\times4$ matrix containing the hamiltonian
contributions [{\it cf.} (2.8)],
$$
{\cal H}=\left[\matrix{2\,{\cal I}m(H_1)&0&iH_3^*&-iH_3\cr
                         0&2\,{\cal I}m(H_2)&-iH_4&iH_4^*\cr
                         iH_4^*&-iH_3&i(H_2^*-H_1)&0\cr
                         -iH_4&iH_3^*&0&i(H_1^*-H_2)}\right]\ , \eqno(3.2)
$$
while the dissipative part $\cal L$ is given in (2.14).

As already mentioned before, the hamiltonian piece in (3.2) 
contains contributions that are not invariant under 
$CPT$ and $T$ transformations; this is
also true for $\cal L$, so that in general dissipation will induce
violations of these discrete symmetries.
It is instructive to explicitly discuss this point in the formalism of (3.1).%
\footnote{$^\ddagger$}{Further discussions on the notion of symmetry
invariance for quantum dynamical semigroups can be found in [43].}

Following the rules of quantum mechanics, any symmetry transformation
can be realized by a unitary or antiunitary operator $U$ acting on the
basis states $|B^0\rangle$, $|\overline{B^0}\rangle$, or alternatively
on the density matrix $\rho$. This induces an action on the
vector $|\rho\rangle$, realized by a $4\times4$ matrix $\cal U$:
$$
|\rho\rangle\rightarrow |\rho'\rangle={\cal U}\, |\rho\rangle\ .
\eqno(3.3)
$$
This transformation leaves the equation (3.1) form-invariant,
{\it i.e.} it is a symmetry of the evolution equation,
provided:
$$
{\cal U}\, {\cal K}\, {\cal U}^{-1}={\cal K}\ ,
\eqno(3.4a)
$$
for unitary transformations, or alternatively
$$
{\cal U}\, {\cal K}\, {\cal U}^{-1}={\cal K}^\dagger\ ,
\eqno(3.4b)
$$
in the case of antiunitary transformations.

For instance, an independent phase change of the two basis vectors,
$$
|B^0\rangle\rightarrow e^{i\phi}|B^0\rangle\ ,\qquad 
|\overline{B^0}\rangle\rightarrow e^{i\bar\phi}|\overline{B^0}\rangle\ ,
\eqno(3.5)
$$
is realized by the $4\times4$ diagonal matrix:
$$
{\cal U}_\phi={\rm diag}\left[\, 1,1,e^{i(\phi-\bar\phi)},
e^{-i(\phi-\bar\phi)}\, \right]\ .
\eqno(3.6)
$$
One easily checks that this transformation is not an invariance of $\cal H$
and hence of (3.1); indeed, the off-diagonal elements $H_3$ and $H_4$
of the effective hamiltonian $H$ do change under (3.5).
As a consequence neither $\sigma$ in (2.12), nor $r_S$ and $r_L$ 
are phase invariant.

In the case of discrete $CPT$, $T$ and $CP$ transformations, one finds
that the corresponding matrix $\cal U$ takes a block diagonal form:
explicitly, one finds:
$$
{\cal U}_{CPT}=\left[\matrix{\sigma_1&0\cr
                             0&\sigma_0}\right]\ ,\qquad
{\cal U}_{T}=\left[\matrix{\sigma_0&0\cr
                             0&\sigma_\varphi}\right]\ ,\qquad
{\cal U}_{CP}=\left[\matrix{\sigma_1&0\cr
                             0&\sigma_\varphi}\right]\ ,\qquad
\eqno(3.7)
$$
where $\sigma_i$, $i=1,2,3$ represent the usual Pauli matrices,
$\sigma_0$ being the identity, while
$$
\sigma_\varphi=\left[\matrix{0&e^{-2i\varphi}\cr
                             e^{2i\varphi}&0}\right]\ ,
\eqno(3.8)
$$
contains the dependence on the phase that defines
the $CP$ transformation of the $B^0$-$\overline{B^0}$ basis states:
$|B^0\rangle\rightarrow e^{i\varphi} |\overline{B^0}\rangle$.

For the hamiltonian contribution $\cal H$,
insertion of the expressions (3.7) in the appropriate invariance condition (3.4)
gives the familiar results. In particular, $CPT$ invariance requires $H_1=H_2$,
or equivalently, recalling the definitions (2.11), (2.12):
$$
r_S=r_L\ ,\qquad \theta=\,0
\eqno(3.9)
$$
while $T$ invariance implies $|H_3|=|H_4|$, {\it i.e.}
$$
|r_S\, r_L|=1\ ,\qquad |\sigma|\equiv {1+\xi\over1-\xi}=1\ ,\qquad \xi=\,0\ .
\eqno(3.10)
$$
Further, notice that imposing both $CPT$ and $T$ invariance, 
hence $CP$ invariance, {\it i.e.} $\theta=\xi=\,0$, readily implies:
$$
r_S=r_L=\sqrt\sigma\ ,\qquad \sigma=e^{-2i\varphi}\ .
\eqno(3.11)
$$

For the dissipative part $\cal L$ of the pseudo hamiltonian $\cal K$
similar invariant conditions can be derived using $(3.4a)$ (the matrix
in (2.14) is in fact hermitian); they impose constraints on the
six dissipative parameters 
$a$, $b$, $c$, $\alpha$, $\beta$ and $\gamma$. More specifically,
invariance under $CPT$-transformations requires the vanishing of
$c$ and $b$, while $T$-invariance imposes the conditions:
$(c-ib)=e^{2i\varphi}(c+ib)$ and 
$(\alpha-\gamma+2i\beta)=e^{4i\varphi}(\alpha-\gamma-2i\beta)$.
Finally, $CP$-invariance requires 
$(c-ib)=-e^{2i\varphi}(c+ib)$ and 
$(\alpha-\gamma+2i\beta)=e^{4i\varphi}(\alpha-\gamma-2i\beta)$.
It should be noticed that all these constraints are perfectly compatible
with the inequalities in (2.13); in other words, complete positivity
does not interfere with the discrete transformations.

Although the available experimental bounds on the $CPT$ and $T$ violating
contributions of the hamiltonian $\cal H$ are not very accurate,
the magnitude of the constants $\theta$ and $\xi$ parametrizing these
violations are expected to be very small.
Despite this, in the discussion that follows we shall try to be as general
as possible, and keep $\theta$ and $\xi$, hence $r_S$ and $r_L$, arbitrary,
unless explicitly stated.

In solving the evolution equation (3.1) however, we shall assume the
dissipative contribution $\cal L$ to be small.
In a phenomenological approach, it is hard to give an apriori
estimate on how large the dissipative effects should be.
However, as already mentioned in the introductory remarks,
a general framework in which dissipation naturally emerges is provided
by the study of subsystems in interaction with large environments.
In such instances, the non-standard effects can be roughly estimated
to be proportional to powers of the typical energy of the system,
while suppressed by inverse powers of the characteristic energy
scale of the environment.[4-7]

In the case of the $B^0$-$\overline{B^0}$ system, these considerations,
together with the general idea that dissipation is induced by
quantum effects at a large, fundamental scale $M_F$, 
lead to predict very small values
for the parameters $a$, $b$, $c$, $\alpha$, $\beta$ and $\gamma$;
using dimensional arguments, an upper bound on the magnitude 
of these parameters can be roughly evaluated to be of order
$m_B^2/M_F$, where $m_B$ is the neutral $B$-meson mass.
If the non-standard effects have a gravitational origin, the scale 
$M_F$ should coincide with the Planck mass $M_P$, and the previous
upper bound would give: $m_B^2/M_P\sim 10^{-18}\ {\rm GeV}$.

These considerations allow treating the dissipative piece $\cal L$
in (3.1) as a perturbation to the hamiltonian contribution $\cal H$.
In order to set up the perturbative expansion, it is useful
to make a change of basis in (3.1), so that the 
hamiltonian piece $\cal H$ becomes diagonal.

As discussed in the previous section,
the effective hamiltonian $H$ can be brought to diagonal form
by the similarity transformation (2.6); the diagonalizing matrix
$V$ in (2.7) can be conveniently decomposed as:
$$
V={\widetilde V}\cdot Q\ ,\qquad 
{\widetilde V}=\left[\matrix{r_S&\phantom{-}r_L\cr
                         1&-1}\right]\ ,\qquad
Q=\left[\matrix{q_S&0\cr
                  0&q_L}\right]\ .
\eqno(3.12)
$$
Since the matrix $Q$ is diagonal, it disappears from the relation (2.6),
that can be equally well be written as:
$$
H={\widetilde V}\, H_0\, {\widetilde V}^{-1}\ .
\eqno(3.13)
$$
When applied to the density matrix $\rho$, this change of basis
induces the transformation 
$\rho\rightarrow \tilde\rho=
{\widetilde V}^{-1}\, \rho\, {\widetilde V}^{\dagger-1}$,
or equivalently:
$$
|\rho\rangle\rightarrow |\tilde\rho\rangle={\cal V}\, |\rho\rangle\ ,
\eqno(3.14)
$$
where
$$
{\cal V}={1\over|r_S+r_L|^2}\
\left[\matrix{1&\phantom{-}|r_L|^2&\phantom{-}r_L^*&\phantom{-}r_L\cr
              1&\phantom{-}|r_S|^2&-r_S^*&-r_S\cr
              1&-r_S^* r_L&-r_S^*&\phantom{-}r_L\cr
              1&-r_S r_L^*&\phantom{-}r_L^*&-r_S}\right]\ .
\eqno(3.15)
$$
Then, the evolution equation (3.1) becomes
$$
{d\over d t}|\tilde\rho(t)\rangle=
\Big[{\cal H}_0+\widetilde{\cal L}\,\Big]\, |\tilde\rho(t)\rangle\ ,
\eqno(3.16)
$$
with 
$$
\widetilde{\cal L}={\cal V}\, {\cal L}\, {\cal V}^{-1}\ ,
\eqno(3.17)
$$
and a diagonal hamiltonian piece:
$$
{\cal H}_0=\left[\matrix{-\gamma_S&0&0&0\cr
                         0&-\gamma_L&0&0\cr
                         0&0&-\Gamma_-&0\cr
                         0&0&0&-\Gamma_+}\right]\ ,
\eqno(3.18)
$$
$$
\Gamma_\pm=\Gamma\pm i\Delta m\ ,\qquad \Delta m=m_L-m_S\ .
\eqno(3.19)
$$
The price to pay for this change of basis is a more complicated expression
for the matrix $\widetilde{\cal L}$, 
representing the dissipative contribution (its explicit form is
collected in Appendix A). However, the entries of $\widetilde{\cal L}$
appear now to be manifestly invariant under the change of phase
in (3.5), and therefore, the same is true for the whole 
evolution equation in (3.16). This is a great advantage since
any solution of (3.16), even an approximate one 
obtained using perturbative methods, 
will result to be manifestly phase-invariant.

We remark that physical
observables, being the result of a trace operation (see below), are
by definition independent from any phase convention. 
Provided a sufficient number
of observables are computed, one can consistently re-express a result
obtained in a given phase convention, into any other phase choice.
However, once a phase choice is adopted and a given approximation used,
conclusions drawn from only a limited number of 
physical observations could lead to incorrect conclusions.
It is therefore always preferable at any step
to work in a phase-independent framework.

The evolution equation in (3.16) can now be solved by iteration, to any order
in the small dissipative parameters
$a$, $b$, $c$, $\alpha$, $\beta$ and $\gamma$. 
Simple manipulations allow to express the vector $|\tilde\rho(t)\rangle$
at time $t$ as the series expansion:
$$
\eqalign{|\tilde\rho(t)\rangle =\, & e^{{\cal H}_0 t}\ |\tilde\rho(0)\rangle
+\int_0^t ds\, e^{{\cal H}_0 (t-s)}\, \widetilde{\cal L}\,
e^{{\cal H}_0 s}\ |\tilde\rho(0)\rangle\cr
&\hskip 1cm +\int_0^t ds_1 \int_0^{s_1} ds_2\, e^{{\cal H}_0 (t-s_1)}\, 
\widetilde{\cal L}\, e^{{\cal H}_0 (s_1-s_2)}\,
\widetilde{\cal L}\, e^{{\cal H}_0 s_2}\ |\tilde\rho(0)\rangle\
+\dots\ .}
\eqno(3.20)
$$
For the considerations that follow, it will be sufficient to 
consider terms that are at most linear in the small parameters, 
thus retaining only the first two terms in the \hbox{r.h.s. of (3.20).}
This suggests a further simplifying assumption, that allows
to write down a more manageable expression for the matrix 
$\widetilde{\cal L}$. 

As it is apparent from its definition in (3.17), the entries of
$\widetilde{\cal L}$ are linear combinations of 
$a$, $b$, $c$, $\alpha$, $\beta$ and $\gamma$, with coefficients that 
depend on $\theta$, $\xi$ and the phase of $\sigma$.
Since $\theta$ and $\xi$ are themselves expected to be small,
they can be safely ignored in those expressions.
In other terms, it appears reasonable to neglect 
$CPT$ and $T$ (hence $CP$) violating effects induced by non-vanishing
$\theta$ and $\xi$, when these violating phenomena mix with the  
dissipative effects.

Within this approximation, the matrix $\widetilde{\cal L}$ takes the form
$$
\widetilde{\cal L}=\left[\matrix{-\widetilde D&\phantom{-}\widetilde D
&-\widetilde C&-\widetilde C^*\cr
                                  \phantom{-}\widetilde D&-\widetilde D
&\phantom{-}\widetilde C&\phantom{-}\widetilde C^*\cr
                                 -\widetilde C^*&\phantom{-}\widetilde C^*
&-\widetilde A&\phantom{-}\widetilde B\cr
                                 -\widetilde C&\phantom{-}\widetilde C
&\phantom{-}\widetilde B^*&-\widetilde A\cr}\right]\ ,
\eqno(3.21)
$$
where
$$
\eqalignno{
&\widetilde A={1\over2}\Big\{[(\alpha+\gamma)+2a
+{\cal R}e[(\alpha-\gamma+2i\beta)\sigma]\Big\}\ ,&(3.22a)\cr
&\widetilde B={1\over2}\Big\{[(\alpha+\gamma)-2a
+{\cal R}e[(\alpha-\gamma+2i\beta)\sigma]
-4i\,{\cal I}m[(c-ib)\sqrt\sigma]\Big\}\ ,&(3.22b)\cr
&\widetilde C={1\over2}\Big\{2\,{\cal R}e[(c-ib)\sqrt\sigma]
+i\,{\cal I}m[(\alpha-\gamma+2i\beta)\sigma]\Big\}\ ,&(3.22c)\cr
&\widetilde D={1\over2}\Big\{(\alpha+\gamma)
-{\cal R}e[(\alpha-\gamma+2i\beta)\sigma]\Big\}\ ,&(3.22d)\cr
}
$$
and $\sigma$ is here the pure phase defined in (3.11).

It is now a matter of a simple computation to perform the integrals
in (3.20) and find the time dependence of the components
$\tilde\rho_1(t)$, $\tilde\rho_2(t)$, $\tilde\rho_3(t)$, $\tilde\rho_4(t)$
of the vector $|\tilde\rho(t)\rangle$, that give the entries
of the density matrix $\tilde\rho(t)$. The explicit expressions are
collected in Appendix B, and will be used in the next sections to
study in detail various neutral $B$-meson decays.

\vskip 1cm

{\bf 4. OBSERVABLES}
\medskip

In the formalism of density matrices, any physical observable of
the neutral $B$-meson system is described by a suitable hermitian
operator $\cal O$. Its evolution in time can be obtained
by taking the trace with the density matrix $\rho(t)$.

Of particular interest are those observables ${\cal O}_f$
that are associated with the decay of a $B$-meson into final
states $f$. In the $|B^0\rangle$, $|\overline{B^0}\rangle$ basis,
${\cal O}_f$ is represented by a $2\times2$ matrix,
$$
{\cal O}_f=\left[\matrix{ {\cal O}_1& {\cal O}_3\cr
                          {\cal O}_4& {\cal O}_2}\right]\ ,
\eqno(4.1)
$$
whose entries can be explicitly written in terms of the two 
independent decay amplitudes
${\cal A}(B^0\rightarrow f)$ and ${\cal A}(\overline{B^0}\rightarrow f)$:
$$
\eqalign{ &{\cal O}_1=|{\cal A}(B^0\rightarrow f)|^2\ ,\cr
          &{\cal O}_2=|{\cal A}(\overline{B^0}\rightarrow f)|^2\ ,}\qquad\qquad
\eqalign{ &{\cal O}_3=\big[{\cal A}(B^0\rightarrow f)\big]^*\, 
{\cal A}(\overline{B^0}\rightarrow f)\ ,\cr
          &{\cal O}_4={\cal A}(B^0\rightarrow f)\, 
\big[{\cal A}(\overline{B^0}\rightarrow f)\big]^*\ .}
\eqno(4.2)
$$
Being a physical quantity, directly accessible to the experiment, its mean value, 
$$
\langle{\cal O}_f\rangle\equiv {\rm Tr}\big[{\cal O}_f\, \rho\big]\ ,
\eqno(4.3)
$$
is however basis independent; it can be computed in 
any specific representation.
In particular, using the transformation $\widetilde V$ introduced in (3.12) 
in the definition (4.3),
the time evolution of $\langle{\cal O}_f\rangle$ can be written as
$$
\langle{\cal O}_f\rangle(t)=
{\rm Tr}\Big[\widetilde{\cal O}_f\, \tilde\rho(t)\Big]=
\sum_{i=1}^4 {\widetilde{\cal O}_i}^*\ \tilde\rho_i(t)\ ,
\eqno(4.4)
$$
where the entries $\widetilde{\cal O}_i$ of the transformed matrix
$\widetilde{\cal O}_f={\widetilde V}^\dagger\, {\cal O}_f\, {\widetilde V}$
have been labelled as in (4.1). This general formula will be repeatedly
used in the coming sections to explicitly compute experimentally relevant
decay rates and asymmetries.

In parametrizing $B^0$-$\overline{B^0}$ decays, it is customary
to introduce the following phase-independent complex quantities:[1, 2]
$$
\lambda_S^f={q_S\over p_S}\, 
{{\cal A}(\overline{B^0}\rightarrow f)\over {\cal A}(B^0\rightarrow f)}\ ,\qquad
\lambda_L^f={q_L\over p_L}\, 
{{\cal A}(\overline{B^0}\rightarrow f)\over {\cal A}(B^0\rightarrow f)}\ .
\eqno(4.5)
$$
They can be used to express the entries of the matrix $\widetilde{\cal O}_f$;
explicitly, one finds:
$$
\widetilde{\cal O}_f=\big|{\cal A}(B^0\rightarrow f)\big|^2\
\left[\matrix{|r_S|^2\,\big|1+\lambda_S^f\big|^2 & 
r_S^* r_L\,\big(1+\lambda_S^f\big)^*\, \big(1-\lambda_L^f\big)\cr
&\cr
              r_S r_L^*\, \big(1+\lambda_S^f\big)\, \big(1-\lambda_L^f\big)^*&
|r_L|^2\,\big|1-\lambda_L^f\big|^2}\right]\ .
\eqno(4.6)
$$
It is sometimes convenient to use instead the parameters:
$$
\mu_S^f={p_S\over q_S}\, 
{{\cal A}(B^0\rightarrow f)\over {\cal A}(\overline{B^0}\rightarrow f)}
={1\over\lambda_S^f}\ ,\qquad
\mu_L^f={p_L\over q_L}\, 
{{\cal A}(B^0\rightarrow f)\over{\cal A}(\overline{B^0}\rightarrow f)}
={1\over\lambda_L^f}\ ;
\eqno(4.7)
$$
one can then rewrite (4.6) as:
$$
\widetilde{\cal O}_f=\big|{\cal A}(\overline{B^0}\rightarrow f)\big|^2\
\left[\matrix{\big|1+\mu_S^f\big|^2 & 
-\big(1+\mu_S^f\big)^*\, \big(1-\mu_L^f\big)\cr
&\cr
            -\big(1+\mu_S^f\big)\, \big(1-\mu_L^f\big)^*&
\big|1-\mu_L^f\big|^2}\right]\ .
\eqno(4.8)
$$
The form (4.6) and (4.8) for the observable $\widetilde{\cal O}_f$
are very general and hold for any final decay state $f$. 
Simplified expressions can however be derived by looking at specific
decay channels. 

Let us first consider decays of the neutral $B$-mesons in semileptonic 
states $h\ell\nu$, where $h$ stands for any allowed charged hadronic state.
We shall be as general as possible, and include in our discussion
violations of the $\Delta B=\Delta Q$ rule.
The amplitudes for the decay of a $B^0$ or $\overline{B^0}$ state
into $h^-\ell^+\nu$ and $h^+\ell^-\bar\nu$ can then be parametrized
in terms of three complex constants, $x_h$, 
$y_h$ and $z_h$, as follows:
$$
\eqalignno{
&{\cal A}(B^0\rightarrow h^-\ell^+\nu)={\cal M}_h (1-y_h)\ , &(4.9a)\cr
&{\cal A}(\overline{B^0}\rightarrow h^+\ell^-\bar\nu)=
{\cal M}_h^* (1+y^*_h)\ , &(4.9b)\cr
&{\cal A}(B^0\rightarrow h^+\ell^-\bar\nu)= z_h\, 
{\cal A}(\overline{B^0}\rightarrow h^+\ell^-\bar\nu)\ , &(4.9c)\cr
&{\cal A}(\overline{B^0}\rightarrow h^-\ell^+\nu)=
x_h\, {\cal A}(B^0\rightarrow h^-\ell^+\nu)\ , &(4.9d)}
$$
where ${\cal M}_h$ is a common factor.
(Note that sometimes,[44, 1, 2] 
the notation $\bar x_h\equiv z_h^*$ is used instead of $z_h$.)
The $\Delta B=\Delta Q$ rule would forbid the decays
$B^0\rightarrow h^+\ell^-\bar\nu$ and 
$\overline{B^0}\rightarrow h^-\ell^+\nu$, so that the parameters $x_h$ and $z_h$
measure the violations of this rule. Instead, $CPT$-invariance 
in the decay process would require
$y_h=\,0$. In view of this, the quantities $x_h$, $y_h$ and $z_h$ 
are expected to be very small. In the computation of semileptonic observables,
it seems therefore justified to keep only first order terms in these parameters,
neglecting terms that contain $x_h$, $y_h$ and $z_h$ multiplied by the dissipative
parameters or the constants $\theta$ and $\xi$, that signal
$CPT$ and $T$ violations in ``mixing''.
Within this approximation, the quantities in (4.5) and (4.7) reduce to
$$
\lambda_S^{h^-}=\lambda_L^{h^-}=\sqrt{\sigma^*}\, x_h\equiv \lambda_h\ ,\qquad
\mu_S^{h^+}=\mu_L^{h^+}=\sqrt{\sigma}\, z_h\equiv \mu_h\ ,
\eqno(4.10)
$$
where $\sigma$ is the pure phase of (3.11).

For hadronic final states, such simplifications are in general not possible.
Nevertheless, to be consistent with the approximation adopted in the
previous section, when the coefficients $\lambda_S$, $\lambda_L$,
$\mu_S$ and $\mu_L$ multiply a dissipative parameter, they should be computed
in the limit of exact $CPT$ and $CP$ symmetries. 
This is particularly relevant when the final state $f$ has a definite 
$CP$-parity $\zeta_f$; within that approximation, one in fact finds:[1]
$$
\lambda_S^f=\lambda_L^f=\zeta_f\ .
\eqno(4.11)
$$
As we will see in the following, this observation will be helpful
in the explicit evaluation of the asymmetries
involving decays into these specific final states.

\vskip 1cm

{\bf 5. SINGLE MESON DECAYS}
\medskip

In this section, we shall study observables connected 
to the time-evolution and decay
of a single, uncorrelated $B^0$-$\overline{B^0}$ system, that can be typically
measured at colliders. Let us indicate with $|\tilde\rho_{B^0}(t)\rangle$,
$|\tilde\rho_{\bar B^0}(t)\rangle$ the time evolution 
according to (3.16) of the states $|\tilde\rho_{B^0}\rangle$,
$|\tilde\rho_{\bar B^0}\rangle$
that represent initial pure $B^0$, $\overline{B^0}$ mesons; their components
can be organized in the two matrices:
$$
\tilde\rho_{B^0}={1\over|r_S+r_L|^2}\left[\matrix{1&1\cr 
                             1&1\cr}\right]\ ,\qquad
\tilde\rho_{\bar B^0}={1\over|r_S+r_L|^2}\left[\matrix{|r_L|^2&-r_S^* r_L\cr 
                           -r_S r_L^*&|r_S|^2\cr}\right]\ .\eqno(5.1)
$$
As discussed in the previous section, 
the probability rate that an initial $|\tilde\rho_{B^0}\rangle$,
$|\tilde\rho_{\bar B^0}\rangle$
state decays at time $t$ into a given final state $f$ described
by the operator ${\cal O}_f$ is given by:
$$
{\cal P}_f(B^0;t)={\rm Tr}\big[\widetilde{\cal O}_f\, \tilde\rho_{B^0}(t)
\big]\ ,\qquad
{\cal P}_f(\overline{B^0};t)={\rm Tr}\big[\widetilde{\cal O}_f\, 
\tilde\rho_{\bar B^0}(t)\big]\ .
\eqno(5.2)
$$
In writing down the explicit form of the probabilities $\cal P$, it is convenient
to introduce the new variable $\tau=t\,\Gamma$; in practice, $\tau$ expresses the
time variable in units of the $B$ lifetime. It is also customary to define
the two combinations:%
\footnote{$^\dagger$}{They are usually called $x_B$ and $y_B$, respectively; we
do not use these labels to avoid confusion with the parameters introduced
in (4.9).}
$$
\omega={\Delta m\over\Gamma}\ ,\qquad\qquad \delta={\Delta\Gamma\over 2\Gamma}\ .
\eqno(5.3)
$$
Although not yet directly measured, $\delta$ is expected to be very small,
$\delta\leq 10^{-2}$. For this reason, in discussing $B^0$-$\overline{B^0}$ 
observables, one often takes the simplified assumption $\delta=0$.
This choice produces however important consequences.
Indeed, in the limit $\Delta\Gamma=\,0$, 
the antihermitian part $\mit\Gamma$ of the
effective hamiltonian $H$ in (2.3) becomes proportional to
the identity. In this case, the similarity transformation (2.6) 
that diagonalizes $H$ turns out to be unitary and the eigenstates
$|B_S\rangle$ and $|B_L\rangle$ in (2.4) become orthogonal.
Since, $\langle B_L| B_S\rangle= p_S\, p_L^*-q_S\, q_L^*$, this implies
the condition
$$
r_S\, r_L^*=1\ ,
\eqno(5.4)
$$
and as a consequence, $|\sigma|=1$, or equivalently $\xi=\,0$, and 
also ${\cal I}m(\theta)=\,0$. Therefore, the condition $\delta=\,0$
implies $T$ conservation in the hamiltonian piece of the
evolution equation, while $CPT$ violation is possible
only if ${\cal R}e(\theta)$ is different from zero.
This fact has clearly important consequences for experimental tests of $CPT$ 
and $T$ invariance in ``mixing''.\hbox{[45, 46]}
In the following, we shall
keep $\delta$ nonvanishing, unless explicitly stated.

We shall first consider observables connected with $B$-meson decays
into semileptonic final states. Using the appropriate observables (4.6), (4.8)
and the results in (4.10) together with those collected in Appendix B, 
expressions for the probabilities (5.2) can be explicitly obtained.
Apart from a common exponential decay factor, they show an oscillatory
behaviour modulated by $\omega$ and further exponential terms 
regulated by $\delta$:
$$
\eqalignno{
{\cal P}_{h^-}(B^0;\tau)={|{\cal M}_h|^2\over2}e^{-\tau}\Bigg\{
&\cos\omega\tau\Bigg[{4\,{\cal R}e(r_S\, r_L^*)\over|r_S+r_L|^2}\,e^{-(A-D)\tau}
-2\,{\cal R}e(y_h)
-{4\,\delta\over\delta^2+\omega^2}{\cal R}e(C)\Bigg]\cr
+&\sin\omega\tau\Bigg[-{4\,{\cal I}m(r_S\, r_L^*)\over
|r_S+r_L|^2}
-2\,{\cal I}m(\lambda_h)+{\cal R}e(B)\Bigg]\cr
+&\cosh\delta\tau\Bigg[{2\big(|r_S|^2+|r_L|^2\big)\over|r_S+r_L|^2}
-2\,{\cal R}e(y_h)
+{4\,\delta\over\delta^2+\omega^2}{\cal R}e(C)\Bigg]\cr
+&\sinh\delta\tau\Bigg[-{2\big(|r_S|^2-|r_L|^2\big)\over|r_S+r_L|^2}
-2\,{\cal R}e(\lambda_h)+{D\over\delta}\Bigg]\Bigg\}\ ,&(5.5a)\cr
&&\cr
&&\cr
{\cal P}_{h^+}(B^0;\tau)={|{\cal M}_h|^2\over2}e^{-\tau}\Bigg\{
&\cos\omega\tau\Bigg[-{4\, e^{-(A-D)\tau}\over|r_S+r_L|^2}
-2\,{\cal R}e(y_h)\Bigg]\cr
+&\sin\omega\tau\Bigg[2\,{\cal I}m(\mu_h)-{\cal R}e(B)
+{4\,\delta\over\delta^2+\omega^2}
{\cal I}m(C)\Bigg]\cr
+&\cosh\delta\tau\Bigg[{4\over|r_S+r_L|^2}
+2\,{\cal R}e(y_h)\Bigg]\cr
+&\sinh\delta\tau\Bigg[-2\,{\cal R}e(\mu_h)+{D\over\delta}
-{4\,\omega\over\delta^2+\omega^2}{\cal I}m(C)\Bigg]\Bigg\}\ ,&(5.5b)\cr
}
$$
where the effects of the non-standard, dissipative phenomena are
controlled by the dimensionless parameters [{\it cf.} (3.22)]
$$
A={\widetilde A\over\Gamma}\ ,\qquad
B={\widetilde B\over\Delta m}\ ,\qquad
C={\widetilde C\over\Gamma}\ ,\qquad
D={\widetilde D\over\Gamma}\ .\eqno(5.6)
$$
The expressions for ${\cal P}_{h^-}(\overline{B^0};\tau)$ and 
${\cal P}_{h^+}(\overline{B^0};\tau)$ are obtained from $(5.5b)$ and $(5.5a)$,
respectively,
by changing the sign of $y_h$ and $C$, and letting 
$r_S\rightarrow 1/r_S$, $r_L\rightarrow 1/r_L$ and
$\lambda_h\leftrightarrow\mu_h$.
One can check that the formulas (5.5) reduce to those presented in [25]
when adopting the particular phase convention for which
$$
r_S\simeq 1+2\,\epsilon_S\ ,\qquad r_L\simeq 1+2\, \epsilon_L\ ,
\eqno(5.7)
$$
and the quantities $\epsilon_S$ and $\epsilon_L$, parametrizing
$CPT$ and $CP$ violation in ``mixing'', are considered small.
Instead, we stress that  the expressions in (5.5) are completely general,
and manifestly independent of any phase-choice.

The probabilities ${\cal P}_h$ are directly accessible and can be studied 
in experiments performed at colliders. A preliminary investigation,
assuming no dissipation, has already been performed at LEP
in the framework of the approximation (5.7) and an upper bound on
the magnitude of ${\cal R}e(\epsilon_S+\epsilon_L)$ has been obtained.[47, 48]
Much more accurate studies can be performed both at the Tevatron
and LHC, where the $\tau$-dependence in (5.5) can actually be observed;
extrapolating from preliminary simulation estimates, 
sensitivities of a few percent
on at least some of the parameters in (5.6) can be reasonably expected.[49, 50]

Instead of analyzing directly the decay rates in (5.5), it might be
more convenient to study specific asymmetries, constructed
by taking suitable combinations of the ${\cal P}_h$'s. 
Two independent simple asymmetries that can 
be formed with the probabilities ${\cal P}_h$ and are likely to
be studied at colliders are:
$$
\eqalignno{
&A_{CPT}(\tau)={   {\cal P}_{h^+}(\overline{B^0};\tau) -
                   {\cal P}_{h^-}(B^0;\tau)\over
                   {\cal P}_{h^+}(\overline{B^0};\tau) +
                   {\cal P}_{h^-}(B^0;\tau)}\ ,  &(5.8a)\cr
&\cr
&A_T(\tau)={ {\cal P}_{h^-}(\overline{B^0};\tau) -
             {\cal P}_{h^+}(B^0;\tau)\over
             {\cal P}_{h^-}(\overline{B^0};\tau)+
			 {\cal P}_{h^+}(B^0;\tau)}\ .&(5.8b)\cr}
$$
The first one is sensible to $CPT$ violating effects,
by comparing the rate difference between the process
$B^0\rightarrow B^0$ and its $CPT$-conjugate
$\overline{B^0}\rightarrow\overline{B^0}$,
while the second asymmetry signals the violation of time reversal by measuring
the rate difference between the process $B^0\rightarrow\overline{B^0}$
and its time-conjugate $\overline{B^0}\rightarrow B^0$.

Unfortunately, these simple observables are not very sensitive to the presence
of the non-standard, dissipative effects. For instance, by using the
expressions in (5.5), one finds:
$$
\eqalign{
A_{CPT}(\tau)=&{\big(|r_S|^2-|r_L|^2\big)\, \sinh\delta\tau
+2\, {\cal I}m(r_S\, r_L^*)\,\sin\omega\tau \over
\big(|r_S|^2+|r_L|^2\big)\, \cosh\delta\tau+
2\, {\cal R}e(r_S\, r_L^*)\, \cos\omega\tau}+2\,{\cal R}e(y_h)\cr
%&\cr
&+{1\over\cosh\delta\tau+\cos\omega\tau}\bigg[
{\cal I}m(\lambda_h-\mu_h)\,\sin\omega\tau 
+{\cal R}e(\lambda_h-\mu_h)\, \sinh\delta\tau \bigg]\cr
&-\bigg({\cosh\delta\tau-\cos\omega\tau\over
\cosh\delta\tau+\cos\omega\tau}\bigg)\,
{4\,\delta \over\delta^2+\omega^2}\ {\cal R}e(C)\ ,\cr
}
\eqno(5.9)
$$
which depends on the dissipative parameters through
${\cal R}e(C)$. However, in (5.9) this parameter appears multiplied 
by a factor $\delta$, so that its contribution to $A_{CPT}(\tau)$
turns out to be suppressed with respect to the other terms;
in particular, the approximation $\delta\approx 0$ would completely 
eliminate the presence of this parameter from (5.9).
The same conclusion holds for the other asymmetry $A_T$.
Therefore, observables of type (5.8) are not suitable
for probing the presence of dissipative effects in neutral $B$-meson physics.

More complicated combinations of
the probability rates ${\cal P}_h$ are needed in order to
isolate the dissipative contributions.
A particularly illuminating example is given by:
$$
A_{\Delta m}(\tau)={ \big[{\cal P}_{h^-}(B^0;\tau)-
{\cal P}_{h^-}(\overline{B^0};\tau)\big]
- \big[{\cal P}_{h^+}(B^0;\tau)- {\cal P}_{h^+}(\overline{B^0};\tau)\big]
\over
{\cal P}_{h^-}(B^0;\tau)+{\cal P}_{h^-}(\overline{B^0};\tau)
+{\cal P}_{h^+}(B^0;\tau)+ {\cal P}_{h^+}(\overline{B^0};\tau)}\ .
\eqno(5.10)
$$
Even in the approximation $\delta=\,0$, this asymmetry retains a
distinctive dependence on the parameters that signal the presence
of dissipation:
$$
A_{\Delta m}(\tau)=\big[{\cal R}e(\theta)\big]^2\, \big(1-\cos\omega\tau\big)
+e^{-A\tau}\, \cos\omega\tau
+\big[{\cal R}e(B) - {\cal I}m(\lambda_h+\mu_h)\big]\, 
\sin\omega\tau\ .
\eqno(5.11)
$$
Assuming the validity of the $\Delta B=\Delta Q$ rule, and neglecting the
quadratic dependence on the $CPT$-violating parameter $\theta$, 
the different time behaviours in $A_{\Delta m}(\tau)$ should allow
a determination of the non-standard parameters $A$ and the real part of $B$.
The actual accuracy of such a determination 
highly depends on the sensitivity of the measure of $A_{\Delta m}$.
Although specific studies have not yet been performed for the form (5.11)
of this asymmetry, 
from the results of available simulations one can nevertheless
expect an accuracy in the determination of the various terms in $A_{\Delta m}$
of about a few percent.[49, 50] This sensitivity is already
enough to give interesting bounds on $A$ and ${\cal R}e(B)$.

Nevertheless, the most simple and accurate 
tests on the extended dynamics in (3.1) 
that can be performed at colliders, using semileptonic decays,
is based on the analysis of time-integrated rates, defined in general as 
$$
{\cal P}_f(B)={1\over\Gamma}\int_0^\infty d\tau\ {\cal P}_f(B;\tau)\ .
\eqno(5.12)
$$
Despite most asymmetries constructed with these integrated
probabilities suffer the same $\delta$-suppression problem mentioned before
in discussing the observables in (5.8), a combination analogous to the one 
in (5.10) is again very sensible to the presence of dissipative
parameters. An explicit evaluation gives, for $\delta=\,0$:
$$
\eqalign{
A_{\Delta m}'={1\over 1+\omega^2}\bigg\{1 +\omega^2\, 
\big[{\cal R}e(\theta)\big]^2
&+\omega\,\Big[{\cal R}e(B)- {\cal I}m(\lambda_h+\mu_h)\Big]\cr
&+ {1\over 1+\omega^2}\Big[(\omega^2-1)A
-2\,\omega^2D\Big]\bigg\}\ .}
\eqno(5.13)
$$
Assuming again the $\Delta B=\Delta Q$ rule and neglecting
quadratic $CPT$-violating effects in ``mixing''
($[{\cal R}e(\theta)]^2=\,0$), 
a measure of $A'_{\Delta m}\, (1+\omega^2)$ not compatible with 1
would clearly signal the presence of non-standard effects.
Such a test can be easily performed at colliders with high accuracy, 
in particular
using the next generation dedicated $B$-meson experiments.

Further useful information on the dissipative parameters in (5.6)
can be obtained by studying observables and asymmetries involving
$B$-meson decays into hadronic final states $f$. The explicit form
of such observables are in general more involved than the ones
containing semileptonic decays, since now one can not resort to
simplifying approximations as in (4.10).
Nevertheless, when the final states $f$ exhibit
a definite $CP$-parity $\zeta_f$, at least within a certain approximation,
manageable expressions for the relevant asymmetries can be obtained.

The typical observable that can be constructed in this case involves the
difference of the two probabilities in (5.2):
$$
A_f(\tau)={  {\cal P}_f(B^0;\tau) -
             {\cal P}_f(\overline{B^0};\tau)\over
             {\cal P}_f(B^0;\tau) +
			 {\cal P}_f(\overline{B^0};\tau)}\ ,
\eqno(5.14)
$$
where, for instance, $f$ can represent $D^+ D^-$, $\pi^+ \pi^-$
or $J/\psi\, K$ final states. These asymmetries will be the target
of very intensive experimental studies: in fact, they offer the possibility
of clean tests of the standard model paradigm for $CP$-violation.
On top of this, they turn out to be very sensible to the determination
of the dissipative parameter $C$ in (5.6).

For simplicity, in presenting the explicit expression for 
the asymmetry $A_f(\tau)$ we shall take $\delta=\,0$
and retain only first order terms in the $CPT$-violating
parameter ${\cal R}e(\theta)$. Within this approximation
the two quantities $\lambda_S^f$ and $\lambda_L^f$ parametrizing
the $B^0$-$\overline{B^0}$ decay into $f$ can be expressed in terms
of a unique amplitude ratio,
$$
\lambda_f=\sqrt{\sigma^*}\
{{\cal A}(\overline{B^0}\rightarrow f)\over {\cal A}(B^0\rightarrow f)}\ ,
\eqno(5.15)
$$
as follows
$$
\lambda_S^f=\lambda_f\,\big[1-{\cal R}e(\theta)\big]\ ,\qquad
\lambda_L^f=\lambda_f\,\big[1+{\cal R}e(\theta)\big]\ .
\eqno(5.16)
$$
Using these results in the computation of the probabilities
in (5.2), one finally obtains:
$$
\eqalign{
A_f(\tau)={2\, {\cal R}e(\lambda_f)\over 1+|\lambda_f|^2}\, {\cal R}e(\theta)
+&{2\,\zeta_f\over \omega}\, {\cal I}m(C)
-\bigg[{2\, {\cal I}m(\lambda_f)\over 1+|\lambda_f|^2}
+{2\,\zeta_f\over \omega}\, {\cal R}e(C)\bigg]\ \sin\omega\tau\cr
+&\bigg[{1-|\lambda_f|^2 \over 1+|\lambda_f|^2}
-{2\,{\cal R}e(\lambda_f)\over 1+|\lambda_f|^2}\,{\cal R}e(\theta)
-{2\,\zeta_f\over \omega}\, {\cal I}m(C)\bigg]\ \cos\omega\tau\ .
}
\eqno(5.17)
$$
The presence of dissipation manifests itself in the modification
of the coefficients of the oscillating terms and 
in the presence of a $\tau$-independent piece.
Neglecting $CPT$-violations in ``mixing'' ($\theta=\,0$)
and assuming a vanishingly small
``direct'' $CP$-violation, {\it i.e.} $|\lambda_f|\approx 1$, 
the different $\tau$-dependence in the three terms of $A_f(\tau)$
allows a determination of both ${\cal R}e(C)$ and ${\cal I}m(C)$,
together with the $CP$-violating parameter ${\cal I}m(\lambda_f)$.
Dedicated $B$-experiments at colliders should be able to
identify the various $\tau$-dependence and therefore provide 
very stringent bounds on the parameter $C$.

On the other hand, integrated asymmetries, constructed from the
probabilities in (5.12), are not very useful in probing the
presence of dissipative effects in neutral $B$-meson decays.
For instance, using the same approximations that lead to (5.17),
the integrated asymmetry $A'_f$ defined as in (5.14) takes the form:
$$
\eqalign{
A'_f={2\, {\cal R}e(\lambda_f)\over 1+|\lambda_f|^2}\, {\cal R}e(\theta)
+{2\,\zeta_f\over \omega}\, {\cal I}m&(C)
-{1\over 1+\omega^2}\ \bigg\{\omega\bigg[{2\, {\cal I}m(\lambda_f)\over 1+|\lambda_f|^2}
+{2\,\zeta_f\over \omega}\, {\cal R}e(C)\bigg]\cr
&-{1-|\lambda_f|^2 \over 1+|\lambda_f|^2}
+{2\, {\cal R}e(\lambda_f)\over 1+|\lambda_f|^2}\, {\cal R}e(\theta)
+{2\,\zeta_f\over \omega}\, {\cal I}m(C)\bigg\}\ .}
\eqno(5.18)
$$
Even for vanishing ${\cal R}e(\theta)$ and $|\lambda_f|=1$, the measure of this
asymmetry alone gives little information on the magnitude of $C$.

\vskip 1cm

{\bf 6. CORRELATED MESON DECAYS}
\medskip

The non-standard, dissipative effects described by the generalized
dynamics (3.1) can be further analyzed in experiments involving
correlated $B^0$-$\overline{B^0}$ mesons, at the so-called $B$-factories.
Indeed, as we shall see, these set-ups appear particularly 
suitable for studying phenomena involving loss of quantum coherence.

In those experiments, 
correlated $B^0$-$\overline{B^0}$ mesons are produced from
the decay of the $\Upsilon(4S)$ resonance.
Since the $\Upsilon$-meson has spin 1, its decay into two spinless bosons produces
an antisymmetric spatial state. In the $\Upsilon$-rest frame, the two neutral 
$B$-mesons are produced flying apart with opposite momenta; 
in the basis $|B^0\rangle$,
$|\overline{B^0}\rangle$, the resulting state can be described by:
$$
|\psi_A\rangle= {1\over\sqrt2}\Big(|B^0,-p\rangle \otimes  
|\overline{B^0},p\rangle -
|\overline{B^0},-p\rangle \otimes  |B^0,p\rangle\Big)\ .\eqno(6.1)
$$
The corresponding density operator $\rho_A$ can be abstractly
expressed in terms of a projector:
$$
\rho_A=|\psi_A\rangle\,\langle \psi_A|\ .
\eqno(6.2)
$$
As in the previous sections, we find it more practical
to use the transformation introduced in (3.12), (3.13) and pass to a different 
representation:
$$
\rho_A\rightarrow \widetilde\rho_A=
\Big[{\widetilde V}^{-1}\otimes{\widetilde V}^{-1}\Big]\
\rho_A\
\Big[{\widetilde V}^{\dagger-1}\otimes{\widetilde V}^{\dagger-1}\Big]\ .
\eqno(6.3)
$$
The $4\times 4$ matrix $\widetilde\rho_A$ can then be explicitly expressed as:
$$
{\widetilde \rho}_A=
{1\over 2\, |r_S+r_L|^2}\Big[{\widetilde P}_1\otimes {\widetilde P}_2\ 
+\ {\widetilde P}_2\otimes {\widetilde P}_1\ 
-\ {\widetilde P}_3\otimes {\widetilde P}_4\ 
-\ {\widetilde P}_4\otimes {\widetilde P}_3\Big]\ ,
\eqno(6.4)
$$
where
$$
{\widetilde P}_1=
    \left(\matrix{1 & 0\cr 0 & 0\cr}\right)\ ,\qquad
{\widetilde P}_2=
    \left(\matrix{0 & 0\cr 0 & 1\cr}\right)\ ,
\eqno(6.5a)
$$
$$
{\widetilde P}_3=
    \left(\matrix{0 & 1\cr 0 & 0\cr}\right)\ ,\qquad
{\widetilde P}_4=
    \left(\matrix{0 & 0\cr 1 & 0\cr}\right)\ .
\eqno(6.5b)
$$

The evolution in time of the density matrix $\widetilde\rho_A$
can be analyzed using the single $B$-meson dynamics $\tilde\rho(t)$
discussed in Sect.3. We shall assume that once produced in
a $\Upsilon$-decay, the two mesons evolve in time each according to the
completely positive map $\gamma_t$ generated by (3.16).%
\footnote{$^\dagger$}{Although other possibilities are in principle
conceivable, this choice is the most natural one:
it assures a consistent single meson dynamics when tracing over the
degrees of freedom of one of the two $B$-mesons.[23]}
This guarantees that the resulting dynamics is completely positive and
of semigroup type. As already mentioned, this assures the positivity 
of the eigenvalues of any physical states at all times: this
is an essential condition for the consistent description
of any physical system.[23, 42]

The density matrix that describes a situation in which
the first $B$-meson has evolved up to proper time $t_1$ 
and the second up to proper time $t_2$ is then given by:
$$
\eqalign{
\widetilde\rho_A(t_1,t_2)&\equiv
\big(\gamma_{t_1}\otimes\gamma_{t_2}\big)\big[\widetilde\rho_A\big]\cr
&={1\over 2\,|r_S+r_L|^2}\Big[\widetilde P_1(t_1)\otimes \widetilde P_2(t_2)\ 
+\ \widetilde P_2(t_1)\otimes \widetilde P_1(t_2)\cr
&\hskip 4cm - \widetilde P_3(t_1)\otimes  \widetilde P_4(t_2)-
\widetilde P_4(t_1)\otimes \widetilde P_3(t_2)\Big]\ ,}
\eqno(6.6)
$$
where $\widetilde P_i(t_1)$ and $\widetilde P_i(t_2)$, $i=1,2,3,4$, 
represent the evolution according to (3.16) of the initial operators 
$\widetilde P_i$ in (6.5), up to the time $t_1$
and $t_2$, respectively (see Appendix C).

The two $B$-mesons that come from a decay of an $\Upsilon$ are 
quantum-mechanically entangled, in a way very similar to that of two spin 1/2
particles coming from a singlet state. As in that case, correlated
measures on the two particles become physically significant.
Indeed, the typical observables that can be studied at $B$-factories 
are double decay rates, {\it i.e.} the probabilities 
${\cal G}(f_1,t_1; f_2,t_2)$ that a meson decays
into a final state $f_1$ at proper time $t_1$, while the other meson
decays into the final state $f_2$ at proper time $t_2$.
They can be computed using:
$$
{\cal G}(f_1,t_1; f_2,t_2)=
\hbox{Tr}\Big[\Big(\widetilde{\cal O}_{f_1}\otimes\widetilde{\cal O}_{f_2}\Big) 
\ \widetilde\rho_A(t_1,t_2)\Big]\ ,
\eqno(6.7)
$$
where $\widetilde{\cal O}_{f_1}$, $\widetilde{\cal O}_{f_2}$ 
represent $2\times2$ hermitian
matrices describing the decay of a single meson into the final
states $f_1$, $f_2$, respectively. They can be identified with one of the
two matrices in (4.6) and (4.8). Let us denote with $\widetilde{\cal O}_f^i$,
$i=1,2,3,4$, the entries of the matrix $\widetilde{\cal O}_f$ when
projected along the operators $\widetilde P_i$:
$$
\widetilde{\cal O}_f=\sum_{i=1}^4 \widetilde{\cal O}^i_f\ \widetilde P_i\ . 
\eqno(6.8)
$$
The observables (6.7) can then be rewritten as
$$
{\cal G}(f_1,t_1; f_2,t_2)=
{1\over 2\,|r_S+r_L|^2}\sum_{i,j=1}^4 
\widetilde{\cal O}_{f_1}^i\, \widetilde{\cal O}_{f_2}^j\ 
{\cal P}_{ij}(t_1,t_2)\ ,
\eqno(6.9)
$$
where the elementary probabilities
$$
\eqalign{
{\cal P}_{ij}(t_1,t_2)
&=\hbox{Tr}\big\{\widetilde P_i\, \widetilde P_1(t_1)\big\}\
\hbox{Tr}\big\{\widetilde P_j\, \widetilde P_2(t_2)\big\}\ +\
\hbox{Tr}\big\{\widetilde P_i\, \widetilde P_2(t_1)\big\}\ 
\hbox{Tr}\big\{\widetilde P_j\, \widetilde P_1(t_2)\big\}\cr
&\qquad
-\hbox{Tr}\big\{\widetilde P_i\,  \widetilde P_3(t_1)\big\}\
\hbox{Tr}\big\{\widetilde P_j\, \widetilde P_4(t_2)\big\}\ -\
\hbox{Tr}\big\{\widetilde P_i\, \widetilde P_4(t_1)\big\}\
\hbox{Tr}\big\{\widetilde P_j\, \widetilde P_3(t_2)\big\}\ ,
}
\eqno(6.10)
$$
can be easily computed to any given accuracy using the solution of the evolution
equation (3.16) collected in Appendix B.

The expressions for the double decay rates in (6.9) can be compared
with the results of the experiment. However, 
much of the analysis at $B$-factories is carried out 
using integrated distributions at fixed time interval $t=t_1-t_2$;
this is a consequence of the short $B$-mesons lifetime
and rapid $B^0$-$\overline{B^0}$ oscillations,
that does not allow a precise enough study of the double time dependence in (6.9).
One is then forced to construct single-time distributions, defined by
$$
\eqalign{
{\mit\Gamma}(f_1,f_2;t)&\equiv\int_0^\infty dt'\, {\cal G}(f_1,t'+t;f_2,t')\cr
&={1\over 2\,|r_S+r_L|^2}\sum_{i,j=1}^4 \widetilde{\cal O}_{f_1}^i\, 
\widetilde{\cal O}_{f_2}^j\ {\mit\Pi}_{ij}(t)\ ,}
\eqno(6.11)
$$
where
$$
{\mit\Pi}_{ij}(t)=\int_0^\infty dt'\  {\cal P}_{ij}(t'+t,t')\ ,
\eqno(6.12)
$$
and $t$ is taken to be positive. For negative $t$, one defines:
$$
{\mit\Gamma}(f_1,f_2;-|t|)=\int_0^\infty dt'\, {\cal G}(f_1,t'-|t|;f_2,t')\ 
\theta(t'-|t|)\ ;
\eqno(6.13)
$$
the presence of the step-function is necessary since the evolution is
of semigroup type, with forward in time propagation, starting from zero
(we can not propagate a $B$-meson before it is created in a $\Upsilon$-decay).
In this case, one easily finds: 
$$
{\mit\Gamma}(f_1,f_2;-|t|)={\mit\Gamma}(f_2,f_1;|t|)\ .
\eqno(6.14)
$$
In the following, we shall always assume: $t\geq0$. 

The explicit form
of the single-time probabilities ${\mit\Pi}_{ij}(t)$ in (6.12) can be
found in Appendix C; inserting these in (6.11), one is then able to
compute the evolution in $t$ of the double decay rate
${\mit\Gamma}(f_1,f_2;t)$ for any specific decay state $f_1$ and $f_2$.
Before presenting the results for some experimentally relevant cases,
let us note that in general the non-standard dynamics in (3.16)
gives results for ${\mit\Gamma}(f_1,f_2;t)$ that are quite different
from those obtained in the usual case 
({\it i.e.} in absence of $\widetilde{\cal L}$).
The most striking difference arises when
the final states coincide $f_1=f_2=f$ and $t$ approaches zero. Due to the
antisymmetry of the initial state $|\psi_A\rangle$ in (6.1), quantum mechanics
predicts a vanishing value for ${\mit\Gamma}(f,f;0)$, while in general
this is not the case for the completely positive dynamics generated
by (3.1). This explains why correlated mesons systems turn out to be the most
natural place to look for effects leading to loss of phase
coherence and dissipation.

The quantities ${\mit\Gamma}(f,f;t)$, with $t$ small,
are therefore very sensitive to the dissipative parameters in (3.22). 
This can be explicitly shown by considering the following combination
of decay rates involving semileptonic final states:
$$
{\cal R}(\tau)={ {\mit\Gamma}(h^+,h^+;\tau)+{\mit\Gamma}(h^-,h^-;\tau)
\over
{\mit\Gamma}(h^+,h^-;\tau)+{\mit\Gamma}(h^-,h^+;\tau)}\ ,
\eqno(6.15)
$$
where again the variable $\tau=t\Gamma$ has been introduced.
Neglecting terms containing $\delta$ times small parameters, one finds:
$$
\eqalign{
{\cal R}(\tau)=&{ \big(1+|r_S|^2 |r_L|^2\big)\, 
\big(\cosh\delta\tau-\cos\omega\tau\big)\over
\big(|r_S|^2 +|r_L|^2\big)\cosh\delta\tau+
2\, {\cal R}e(r_S r_L^*)\, \cos\omega\tau}\cr
&+{2\over(\cos\omega\tau +1)^2}\,\bigg[A(1+\tau)\cos\omega\tau
-\big(\omega\cos\omega\tau+\sin\omega\tau\big)\,
{\cal R}e\bigg({B\over1-i\omega}\bigg)\bigg]\ .}
\eqno(6.16)
$$
As $\tau$ approaches zero, the first term becomes vanishingly small,
while the surviving pieces are all proportional to the dissipative parameters,
so that:
$$
{\cal R}(0)={1\over2}\bigg\{A+{\omega\over1+
\omega^2}\big[\omega\, {\cal I}m(B)-{\cal R}e(B)\big]\bigg\}\ .
\eqno(6.17)
$$
A non vanishing result in the measure of this quantity would clearly
signal the presence of dissipation. However, to obtain more detailed
information on the magnitude of the non-standard parameters,
one has to study the full $\tau$-dependence in (6.16), and further 
analyze the behaviour of other observables that can be constructed out of the
double \hbox{probabilities $\mit\Gamma$}.

For instance, one can reverse the sign in the numerator of (6.15)
and study the asymmetry:
$$
{\cal A}_T(\tau)={{\mit\Gamma}(h^+,h^+;\tau) - {\mit\Gamma}(h^-,h^-;\tau)
\over
{\mit\Gamma}(h^+,h^-;\tau)+{\mit\Gamma}(h^-,h^+;\tau)}\ .
\eqno(6.18)
$$
Unfortunately, the same phenomenon already noticed in the case
of single meson asymmetries occurs also here: 
the dependence on the dissipative parameters
is always suppressed by a factor $\delta$, so that
it becomes extremely difficult
to extract information about the non-standard effects
from these observables.

The situation is different for
the asymmetry ${\cal A}_{\Delta m}(\tau)$, analogous to the one in (5.10)
for single-meson systems:
$$
{\cal A}_{\Delta m}(\tau)={ \big[{\mit\Gamma}(h^+,h^-;\tau)
+{\mit\Gamma}(h^-,h^+;\tau)\big] - \big[{\mit\Gamma}(h^+,h^+;\tau)
+{\mit\Gamma}(h^-,h^-;\tau)\big]
\over
{\mit\Gamma}(h^+,h^+;\tau) +{\mit\Gamma}(h^-,h^-;\tau)
+{\mit\Gamma}(h^+,h^-;\tau) + {\mit\Gamma}(h^-,h^+;\tau)}\ .
\eqno(6.19)
$$
In fact, even in the case $\delta=\,0$, this asymmetry has a very distinctive
dependence on the non-standard parameters:
$$
{\cal A}_{\Delta m}(\tau)=
{e^{-A\tau}\over 1+A}\, \cos\omega\tau +
\big[{\cal R}e(\theta)\big]^2\big(1-\cos\omega\tau\big)
+\big(\omega\cos\omega\tau+\sin\omega\tau\big)\,
{\cal R}e\bigg({B\over1-i\omega}\bigg)\ .
\eqno(6.20)
$$
Thanks to the different time behaviour, an experimental study 
of this observable should allow to extract precise information
on $A$ and the combination $\omega\,{\cal I}m(B)-{\cal R}e(B)$;
the analysis of ${\cal R}(\tau)$ in (6.16) should then provide
an estimate on the parameter $D$.

The asymmetry in (6.19) is presently under intensive study
at $B$-factories:[51-53] 
it is used to obtain a precise determination of the mass
difference $\Delta m$, since in absence of dissipation:
${\cal A}_{\Delta m}(\tau)=\cos\omega\tau$. Although an 
accurate analysis is certainly needed in order to estimate precisely
the sensitivity of those experiments 
to the different $\tau$-dependent functions
in (6.20), on the basis of estimated data acquisition and relative
errors, one can reasonably expect that the actual measured data will at the end
constrain the dissipative constants $A$ and $\omega\,{\cal I}m(B)-{\cal R}e(B)$
to a few percent level. 
This would allow to establish significant limits on the presence 
of dissipative effects in the $B^0$-$\overline{B^0}$ system.

To get information on the magnitude of the fourth parameter $C$ in (5.6),
one has to study observables that involve also hadronic decay channels,
in particular those with definite $CP$-parity. A typical asymmetry
that is measured at $B$-factories involves double decay rates
${\mit\Gamma}(f_1,f_2;\tau)$ in which one of the final states is
semileptonic and the other is hadronic:
$$
{\cal A}_f(\tau)={ {\mit\Gamma}(h^+,f;\tau)-{\mit\Gamma}(h^-,f;\tau)
\over
{\mit\Gamma}(h^+,f;\tau)+{\mit\Gamma}(h^-,f;\tau)}\ .
\eqno(6.21)
$$
Since the two mesons are quantum mechanically correlated, 
neglecting $\Delta B=\Delta Q$ violating effects, the
semileptonic decay effectively ``tags'' the flavour of the meson that decays
into $f$. In view of this, one usually consider the asymmetry in
(6.21) and the one in (5.14) for single meson decays
as the same observable;[1, 2] for instance, when  $f=J/\psi\, K_S$ both
asymmetries can be directly expressible in terms of one of the
angles parametrizing $CP$ violation in the standard model.
The situation radically changes 
in presence of dissipation;
as already stressed, the form of the double decay rates
${\mit\Gamma}(f_1,f_2;\tau)$ turns out to be very different
from that predicted by ordinary quantum mechanics and this
makes the asymmetry (6.21) distinct from that in (5.14).

In presenting the explicit expression of ${\cal A}_f(\tau)$,
one can set $\delta=\,0$, since this choice does not affect
the dependence on the dissipative parameters. Further,
we shall assume that $CPT$-violation in ``mixing'' be small,
and therefore neglect terms
containing powers of the  parameter ${\cal R}e(\theta)$
higher then one. Within these approximations, one explicitly finds:
$$
\eqalign{
{\cal A}_f(\tau)=
&{2\, {\cal R}e(\lambda_f)\over 1+|\lambda_f|^2}\, {\cal R}e(\theta)
+\zeta_f\, {\cal R}e\big(\lambda_h-\mu_h+2\,\zeta_f\, y_h\big)
-{2\,\zeta_f\over \omega}\, {\cal I}m(C)\cr
&\hskip 1cm+\bigg[{2\, {\cal I}m(\lambda_f)\over 1+|\lambda_f|^2}\,
-{2\,\zeta_f\over \omega}\, {\cal I}m\bigg(
{2i+\omega\over2+i\omega}\, C\bigg)\bigg]\ \sin\omega\tau\cr
&\hskip 1cm +\bigg[{1-|\lambda_f|^2 \over 1+|\lambda_f|^2}
-{2\, {\cal R}e(\lambda_f)\over 1+|\lambda_f|^2}\,{\cal R}e(\theta)
-{2\,\zeta_f\over \omega}\,{\cal R}e\bigg(
{2i+\omega\over2+i\omega}\, C\bigg)\bigg]\ \cos\omega\tau\ ,}
\eqno(6.22)
$$
where the decay parameter $\lambda_f$ is defined in (5.15), while
$\zeta_f$ represents again the intrinsic $CP$ parity of the state $f$.

Assuming the validity of the $\Delta B=\Delta Q$ rule and neglecting
$CPT$-violating effects in ``mixing''($\theta=\,0$) as well as
``direct'' $CP$-violations ($|\lambda_f|=1$), a fit of 
(6.22) with experimental data allows the determination of both
${\cal R}e(C)$ and ${\cal I}m(C)$, together with ${\cal I}m(\lambda_f)$.
The asymmetry ${\cal A}_f$, in particular for the final state $f=J/\psi\, K_S$,
will be measured with increasing accuracy at $B$-factories,
so that a high sensitivity on the different $\tau$ dependences in (6.22)
is expected.[51] Therefore, the study of ${\cal A}_f$ could result in
one of the best tests on the presence of dissipative effects
in $B$ physics.

As remarked before, the form of ${\cal A}_f(\tau)$ in (6.22)
is valid only for $\tau\geq0$. For negative times, one has to use
the an analogous expression, obtained from the definition in (6.21)
by exchanging the positions of the semileptonic and $f$ final states
in the probabilities $\mit\Gamma$. Explicit computations gives:
$$
\eqalign{
{\cal A}_f(-|\tau|)=
&{2\, {\cal R}e(\lambda_f)\over 1+|\lambda_f|^2}\, {\cal R}e(\theta)
+\zeta_f\, {\cal R}e\big(\lambda_h-\mu_h+2\,\zeta_f\, y_h\big)
-{2\,\zeta_f\over \omega}\, {\cal R}e\bigg(
{2i+\omega\over2+i\omega}\, C\bigg)\cr
&\hskip 1cm -\bigg[{2\, {\cal I}m(\lambda_f)\over 1+|\lambda_f|^2}\,
+{2\,\zeta_f\over \omega}\, {\cal R}e(C)\bigg]\ \sin\omega|\tau|\cr
&\hskip 1cm +\bigg[{1-|\lambda_f|^2 \over 1+|\lambda_f|^2}
-{2\, {\cal R}e(\lambda_f)\over 1+|\lambda_f|^2}\, {\cal R}e(\theta)
-{2\,\zeta_f\over \omega}\,{\cal I}m(C)\bigg]\ \cos\omega|\tau|\ .}
\eqno(6.23)
$$
Note that the two expressions (6.22) and (6.23) coincide at $\tau=\,0$,
while in absence of dissipation ($C=\,0$), (6.23) can be obtained
from (6.22) by letting $\tau\rightarrow -\tau$; the fact that this
no longer true for $C\neq0$ is a clear sign 
of presence of irreversibility.

The observables involving correlated $B$-mesons
that have been discussed so far are accessible at the so-called asymmetric
$B$-factories, where the $\Upsilon$-decays take place in a boosted
reference frame with respect to the laboratory. This allows 
the subsequent decays of the two neutral $B$ mesons to be physically separated
and therefore a good determination of the time-difference $\tau$.
In symmetric $B$-factories, this measure is impossible:
the two mesons decay too quickly to allow a complete reconstruction
of the events. One has then to resort to time-independent probabilities,
obtained by further integrating the rates ${\mit\Gamma}(f_1, f_2;t)$
in (6.11):
$$
{\mit\Gamma}(f_1,f_2)=\int_0^\infty d t\
{\mit\Gamma}(f_1,f_2;t)\ .
\eqno(6.24)
$$

The asymmetries that can be constructed out of these quantities 
are not very sensitive to the non-standard parameters, because of the
suppression by factors $\delta$. Nevertheless, one can still form 
useful observables involving semileptonic final states;
these are the total $B^0$-$\overline{B^0}$ mixing
probability:
$$
\chi_B={ {\mit\Gamma}(h^+,h^+)+{\mit\Gamma}(h^-,h^-)\over
{\mit\Gamma}(h^+,h^+)+{\mit\Gamma}(h^-,h^-)+
{\mit\Gamma}(h^+,h^-)+{\mit\Gamma}(h^-,h^+)}\ ,
\eqno(6.25)
$$
and the ratio of the total, same-sign to opposite-sign semileptonic 
rates:
$$
R_B={ {\mit\Gamma}(h^+,h^+)+{\mit\Gamma}(h^-,h^-)\over
{\mit\Gamma}(h^+,h^-)+{\mit\Gamma}(h^-,h^+)}\ .
\eqno(6.26)
$$
Both observables have a quadratic dependence on $\delta$;
the approximation $\delta=\,0$ is therefore very accurate.
With this choice, one explicitly finds:
$$
\eqalignno{
&\chi_B={\omega^2\over 2(1+\omega^2)}
\bigg\{1-\big[{\cal R}e(\theta)\big]^2+{2\over\omega^2(1+\omega^2)}\, X\bigg\}
\ , &(6.27a)\cr
&R_B={\omega^2\,\big(1-[{\cal R}e(\theta)]^2\big)\over 
2+\omega^2\, \big(1+[{\cal R}e(\theta)]^2\big)}
+{4\over(2+\omega^2)^2}\, X
\ , &(6.27b)}
$$
where the dependence on the non-standard, dissipative parameters
occurs via the combination:
$$
X=A+\omega^2\, D+\omega\big[\omega\, {\cal I}m(B) - {\cal R}e(B)\big]\ .
\eqno(6.28)
$$
Independent measures of these two quantities would provide
a way to estimate both $[{\cal R}e(\theta)]^2$ and $X$,
and therefore give limits on both dissipative and $CPT$-violating effects.
Unfortunately, both observables are not very well known;
$\chi_B$ is the better determined parameter
and the most recent data%
\footnote{$^\dagger$}{The determination of $\chi_B$ in [54] actually involves
so-called semileptonic-type decays [36] rather than pure semileptonic 
final states; one can check that the form $(6.27a)$ for the observable
$\chi_B$ is valid also in that case.}
give: $\chi_B=0.198\pm0.019$.[54]
Assuming $[{\cal R}e(\theta)]^2$ to be negligible and using the world
average for $\omega$,[55] from $(6.27a)$ one gets the estimate:
$X=(5.7\pm4.8)\times 10^{-2}$.%
\footnote{$^\ddagger$}{When $a=\,0$, the inequalities (2.13) 
further imply $\alpha=\gamma$ and $b=c=\beta=\,0$, or equivalently
$A=B=D$, $C=\,0$. In this simplified case, an estimate on the variable $X$
translates into a corresponding one for the surviving dissipative
parameter $\alpha$; in particular, from the above quoted value for $X$, one
has: $\alpha=(4.5\pm3.8)\times 10^{-14}\ {\rm GeV}$.}
The accuracy on the determination 
of $X$ will be greatly improved when also the measurements at 
the asymmetric $B$-factories will be available.
Indeed, preliminary estimates of
about one percent sensitivity to $R_B$ have been reported to be attainable
in [38], while a conservative estimate of two percent accuracy in the measure
of $\chi_B$ is indicated in [40]. If confirmed by the actual data, these
sensitivities will allow a determination of the combination $X$ with
about a few percent accuracy, providing an interesting test
on the presence of non-standard effects in $B$-physics.

%\vfill\eject
\vskip 1cm

{\bf 7. DISCUSSION}
\medskip

The description of open quantum systems in terms of quantum dynamical
semigroups allows a general and consistent approach to the study
of physical phenomena leading to irreversibility and dissipation.
When applied to the analysis of the propagation and decay of
neutral meson systems, it gives precise predictions
on the behaviour of relevant physical observables:
the new, dissipative phenomena manifest themselves through
a set of phenomenological parameters,
$a$, $b$, $c$, $\alpha$, $\beta$ and $\gamma$,
whose presence can be experimentally probed.

Indeed, as discussed at length in the previous sections, various observables
involving $B^0$-$\overline{B^0}$ decays can be identified as being
particularly sensitive to the new, non-standard effects.
These observables will be measured with great accuracy in the new
generation of dedicated $B$-experiments, both at colliders
(CDF-II, HERA-B, BTeV, LHC-$b$) and at $B$-factories
(BaBar, Belle, CLEO-III), so that stringent bounds
on the dissipative effects can be expected in the future.

From the experimental point of view, the actual visibility of these
effects clearly depends on the magnitude
of the parameters $a$, $b$, $c$, $\alpha$, $\beta$ and $\gamma$.
A pure phenomenological approach can not provide
such information. However, in the framework of open quantum systems,
the neutral $B$-mesons are described as
subsystems in interaction with an environment.
In such instances, the effects of irreversibility and dissipation
can be roughly estimated to be proportional to the square
of the $B$-meson mass divided by the characteristic energy
scale of the environment. Assimilating this scale to the
Planck mass results in a very small estimate for the
magnitude of the dissipative parameters, roughly of order $10^{-18}\ {\rm GeV}$.
However, the sophistication of the experiments mentioned before
is so high, that the sensitivity needed to provide useful constraints on
such tiny effects should be reached in just a few years of data taking.[49-53]

Being based on the general theory of open systems,
the above estimate on the magnitude of the non-standard, dissipative effects
is rather robust and quite independent from the details of the actual 
dynamics that drives the interaction between subsystem and environment.
Nevertheless, it has been recently questioned,[56] on the basis
of the similarity of a simplified version of the evolution equation
in (3.1) with those describing the dynamical reduction of the wave-packet.
We point out that this analogy is only superficial: the physical
effects leading to dissipation in open systems are clearly distinct from those
advocated as responsible for the reduction process; and indeed, the quantum
dynamical semigroup generated by (3.1) can not describe dynamical
reduction phenomena. 

The arguments of [56] have been further used in [57] to support the claim
that non-linear evolution equations should be used to analyze
dissipative effects in neutral meson systems,
although also there dynamics of the form (2.2) 
are nevertheless adopted at the end.
In the framework of open systems, the situation can be easily clarified.

The dynamics of a small system $S$ in interaction with a 
large environment $E$ is in general very complex and can not be described 
in terms of evolution equations that are local in time: 
possible initial correlations 
and the continuous exchange of energy as well as entropy between the
$S$ and $E$ produce memory effects and non-linear phenomena.
Nevertheless, when the typical time scale in the evolution 
of the subsystem $S$ is much larger than the characteristic time 
correlations in the environment, the subdynamics simplifies
and a mathematically precise description
in terms of quantum dynamical semigroups naturally emerges.[4-6]

This description is very general and 
is applicable to all physical situations for which
the interaction between $S$ and $E$ can be considered to be
weak and for times for which non-linear disturbances due to possible
initial correlations have disappeared.[7]
These are precisely the conditions that are expected to be fulfilled
in neutral meson systems: the characteristic time correlations
in the environment, induced by the fundamental 
({\it e.g.} gravitational or ``stringy'')
dynamics, is certainly much smaller than the neutral meson lifetime,
and the interaction between mesons and environment is for sure weak
(its effects have not yet been detected). 
Furthermore, this open system
paradigm automatically assures the fulfillment of basic physical properties,
as forward in time composition and entropy increase (irreversibility).
Therefore, the physical motivations for adopting a quantum dynamical semigroup 
description of the extended neutral-meson effective dynamics appear to be
rather compelling and general.

As a final remark, let us point out that once effective dynamics
generated by equations of the form (2.2) are accepted,
the condition of complete positivity is then absolutely necessary
for a physically consistent description of the neutral meson system.
Indeed, time evolutions that do not satisfy this property
unavoidably gives unphysical results.[42]
This can be easily shown by examining the sign of the eigenvalues of
of the density matrix $\widetilde\rho_A(t,t)$ representing a correlated
meson state. 
In fact, one of the properties that any density matrix needs to satisfy is 
that its eigenvalues be non-negative, for all times; without this basic
requirement, its standard probability interpretation
would be meaningless.

As an example, following [28, 19, 20],
let us consider a dynamics for $\widetilde\rho_A$ generated by single-meson
evolution equation of the form (3.1),
where $\cal L$ is as in (2.14), but with vanishing $a$, $b$ and $c$. 
The conditions of complete positivity in (2.13) are then clearly violated
(unless $\alpha=\gamma$ and $\beta=\,0$). For ${\cal H}=\,0$,
the eigenvalues $\lambda_i(t)$, $i=1,2,3,4$, of the $4\times4$ matrix 
$\widetilde\rho_A(t,t)$ can be explicitly obtained; 
being physical quantities, they can be computed in any phase
convention, and in particular in the one 
for which the dissipative parameters in (3.22)
take the simplified form
$\widetilde A=\widetilde B=\alpha$, $\widetilde C=i\beta$
and $\widetilde D=\gamma$. Then, one has:
$$
\eqalign{
&\lambda_{1,2}(t)=2\pm\Big\{\big[E_+(t)\big]^2 + \big[E_-(t)\big]^2 
+2 \big[F(t)\big]^2\Big\}\ ,\cr
&\lambda_{3,4}(t)=\pm\big[E_+(t)+E_-(t)\big]\,
\Big\{ \big[E_+(t)-E_-(t)\big]^2 + 4 \big[F(t)\big]^2\Big\}^{1/2}\ ,}
\eqno(7.1)
$$
where
$$
\eqalign{
&E_\pm(t)={1\over \nu_+-\nu_-}\left[(\nu_+ +2\alpha)\,e^{\nu_\pm\,t}-
(\nu_-+2\alpha)\,e^{\nu_\mp\,t}\right]\ ,\cr
&F(t)={2\beta\over \nu_+ -\nu_-}
\left[e^{\nu_-\,t}-e^{\nu_+\,t}\right]\ ,}
\eqno(7.2)
$$
and
$\nu_\pm=-(\alpha+\gamma)\pm\sqrt{(\alpha-\gamma)^2+4\beta^2}$
are both negative due to the simple positivity condition 
$\alpha\gamma\geq\beta^2$. As clear from the previous formulas,
for $t\neq0$ two of the eigenvalues 
of $\widetilde\rho_A(t,t)$ are always negative,
signaling the presence of unphysical ``negative
probabilities''. When complete positivity is enforced though, one has
$\nu_+ = \nu_-$ and all eigenvalues in (7.1) remain non-negative for all times.

These results offer a further motivation for studying generalized
dynamics of the form (3.1) at meson factories.
These set-ups are in fact high-performance
quantum interferometers: at least in principle,  
they can clarify from the experimental point of view the role of
the condition of complete positivity in the time evolution
of correlated mesons.[42]
As shown by the simple example above, this is not just a
mere technical question: it is crucial to our physical understanding
of the quantum dynamics of open systems.

\vskip 2cm

%\vfill\eject

{\bf APPENDIX A}
\medskip

As explained in the text, in order to find explicit solutions for the
evolution equation (3.1), it is convenient to make a change of basis
as described in (3.14). The resulting dynamical equation
has a simplified hamiltonian part ${\cal H}_0$, while the dissipative
piece becomes more involved:
$\widetilde{\cal L}={\cal V}\, {\cal L}\, {\cal V}^{-1}$.
Its entries can be expressed as linear combinations of the
dissipative parameters $a$, $b$, $c$, $\alpha$, $\beta$ and $\gamma$.
One explicitly finds:
$$
\widetilde{\cal L}={1\over|r_S+r_L|^2}\
\left[\matrix{\Lambda&\Sigma&\Delta&\Delta^*\cr
               \Xi&\Lambda&\Phi&\Phi^*\cr
                -\Phi^*&-\Delta^*&\Omega&\Theta\cr
                 -\Phi&-\Delta&\Theta^*&\Omega}\right]\ ,
\eqno(A.1)
$$
where
$$
\eqalign{
&\Lambda=a\big(|r_S|^2-1\big)\big(|r_L|^2-1\big)
+2\,{\cal R}e\Big\{(c-ib)\Big[\big(|r_L|^2-1\big)r_S
-\big(|r_S|^2-1\big)r_L\Big]\Big\}\cr
&\hskip 5cm -2(\alpha+\gamma)\, {\cal R}e\big(r_S r_L^*\big)
+2\, {\cal R}e\Big\{(\alpha-\gamma+2i\beta)r_S r_L\Big\}\ ,\cr 
&\cr
&\Sigma=a\big(|r_L|^2-1\big)^2
-4\,{\cal R}e\Big\{(c-ib)\big(|r_L|^2-1\big) r_L\Big\}
+2(\alpha+\gamma)|r_L|^2\cr
&\hskip 8cm -2\, {\cal R}e\Big\{(\alpha-\gamma+2i\beta)r_L^2\Big\}\ ,\cr
&\Xi=a\big(|r_S|^2-1\big)^2
+4\,{\cal R}e\Big\{(c-ib)\big(|r_S|^2-1\big) r_S\Big\}
+2(\alpha+\gamma)|r_S|^2\cr
&\hskip 8cm -2\, {\cal R}e\Big\{(\alpha-\gamma+2i\beta)r_S^2\Big\}\ ,\cr
&\cr
&\Delta=a\big(|r_L|^2-1\big)\big(r_S r_L^*+1\big)
+(\alpha+\gamma)\big(r_S-r_L\big)r_L^*
+(c-ib)\big(r_S-r_L-2r_S |r_L|^2)\cr
&\hskip 1cm +(c+ib)\big[\big(r_L-r_S\big)r_L^*-2\big]r_L^*
+(\alpha-\gamma-2i\beta)r_L^*{}^2
-(\alpha-\gamma+2i\beta)r_S r_L\ ,\cr
&\cr
}
$$

\vfill\eject

$$
\eqalign{
&\Phi=a\big(|r_S|^2-1\big)\big(r_S r_L^*+1\big)
+(\alpha+\gamma)\big(r_L^*-r_S^*\big)r_S
+(c-ib)\big[\big(r_L^*-r_S^*\big)r_S+2\big]r_S\cr
&\hskip 2cm +(c+ib)\big(r_S^*-r_L^*+2|r_S|^2 r_L^*\big)
-(\alpha-\gamma-2i\beta)r_S^* r_L^*
+(\alpha-\gamma+2i\beta)r_S^2\ ,\cr
&\cr
&\Theta=-a\big(r_S^* r_L+1\big)^2
-2(c-ib)\big(r_S^* r_L+1\big) r_L
+2(c+ib)\big(r_S^* r_L+1\big) r_S^*
+2(\alpha+\gamma)r_S^* r_L\cr
&\hskip 7cm +(\alpha-\gamma-2i\beta)r_S^*{}^2
+(\alpha-\gamma+2i\beta)r_L^2\ ,\cr
&\cr
&\Omega=-a\big|r_S r_L^*+1\big|^2
+2\,{\cal R}e\Big\{(c-ib)\Big[\big(|r_S|^2-1\big)r_L
-\big(|r_L|^2-1\big)r_S\Big]\Big\}\cr
&\hskip 4cm -(\alpha+\gamma)\big(|r_S|^2+|r_L|^2\big)
-2\, {\cal R}e\Big\{(\alpha-\gamma+2i\beta)r_S r_L\Big\}\ .\cr
}
$$
\null

\noindent
Recalling the discussion in Section 3, an independent phase change of the
basis vectors, $|B^0\rangle\rightarrow e^{i\phi}\, |B^0\rangle$,
$|\overline{B^0}\rangle\rightarrow e^{i\bar\phi}\, |\overline{B^0}\rangle$,
induces a transformation on the entries of the matrix $\cal L$ in (2.14),
realized by the operator ${\cal U}_\phi$ in (3.6); explicitly, one finds:
$$
(c-ib)\rightarrow e^{-i(\phi-\bar\phi)}\, (c-ib)\ ,\qquad
(\alpha-\gamma-2i\beta)\rightarrow e^{2i(\phi-\bar\phi)}\, 
(\alpha-\gamma-2i\beta)\ ,
\eqno(A.2)
$$
while $a$ and $\alpha+\gamma$ remain unchanged. Similarly, also the two
ratios $r_S$ and $r_L$ in (2.9) change, according to the rule:
$$
r_S\rightarrow e^{i(\phi-\bar\phi)}\ r_S\ ,\qquad
r_L\rightarrow e^{i(\phi-\bar\phi)}\ r_L\ .
\eqno(A.3)
$$
As a result, the entries of the matrix $\widetilde{\cal L}$ listed above
are manifestly rephasing invariant.

\vskip 2cm
%\vfill\eject

{\bf APPENDIX B}
\medskip

In order to study the time evolution of $B^0$-$\overline{B^0}$ observables, 
one has to solve the evolution equation (3.16),
$$
{d\over d t}|\tilde\rho(t)\rangle=
\Big[{\cal H}_0+\widetilde{\cal L}\,\Big]\, |\tilde\rho(t)\rangle\ ,
\eqno(B.1)
$$
with $\widetilde{\cal L}$ as in (3.21), 
for a given initial state $|\tilde\rho(0)\rangle$.
In other terms, one has to
compute the entries of the $4\times 4$ evolution matrix $M_{ij}(t)$,
which gives the components $\tilde\rho_1(t)$, $\tilde\rho_2(t)$,
$\tilde\rho_3(t)$, $\tilde\rho_4(t)$ of the state vector $|\tilde\rho(t)\rangle$
at time $t$, in terms of the initial ones at $t=\,0$:
$$
\tilde\rho_i(t)=\sum_{j=1}^4 M_{ij}(t)\, \tilde\rho_j(0)\ ,\qquad i=1,2,3,4\ .
\eqno(B.2)
$$
As explained in the text, it is sufficient to write down an
approximate expression for $M_{ij}(t)$ that contains contributions
up to first order in the dissipative parameters (3.22) appearing in
$\widetilde{\cal L}$. 
The expansion of $M_{ij}(t)$ within this approximation can be conveniently
organized as the sum of two contributions:
$$
\tilde\rho_i(t)\simeq\sum_{j=1}^4
\Big[ M^{(0)}_{ij}(t)\ +\ M^{(1)}_{ij}(t)\Big]\, \tilde\rho_j(0)\ .
\eqno(B.3)
$$
The matrix $M^{(0)}$ has only diagonal
non-vanishing terms:
$$
\eqalign{
&M^{(0)}_{11}(t)=e^{-\gamma_S t}\ ,\cr
&M^{(0)}_{22}(t)=e^{-\gamma_L t}\ ,\cr}\qquad\qquad
\eqalign{
&M^{(0)}_{33}(t)=e^{-(\Gamma_- + \widetilde A-\widetilde D)t}\ ,\cr
&M^{(0)}_{44}(t)=e^{-(\Gamma_+ +\widetilde A-\widetilde D)t}\ .}
\eqno(B.4)
$$
The entries of $M^{(1)}$ take instead the following explicit expression:

$$
\eqalign{
&M^{(1)}_{11}(t)=0 \phantom{\gamma\over\Delta\Gamma}\cr
&M^{(1)}_{13}(t)={2\widetilde C\over\Delta\Gamma_+}\Big(
                  e^{-\gamma_S t}- e^{-\Gamma_- t}\Big)\ \cr
&\cr
&M^{(1)}_{21}(t)={\widetilde D\over\Delta\Gamma}\Big(
                  e^{-\gamma_L t}-e^{-\gamma_S t}\Big)\ \cr
&M^{(1)}_{23}(t)={2\widetilde C\over\Delta\Gamma_-}\Big(
                         e^{-\gamma_L t}-e^{-\Gamma_-  t}\Big)\ \cr
&\cr
&M^{(1)}_{31}(t)={2\widetilde C^*\over\Delta\Gamma_+}\Big(e^{-\gamma_S t}
                                 -e^{-\Gamma_-  t}\Big)\ \ \cr
&M^{(1)}_{33}(t)=0 \phantom{iB\over2\Delta m}\cr
&\cr
&M^{(1)}_{41}(t)={2\widetilde C\over\Delta\Gamma_-}\Big(e^{-\gamma_S t}
                               -e^{-\Gamma_+ t}\Big)\ \cr
&M^{(1)}_{43}(t)={i\widetilde B^*\over2\Delta m}\,
                  \Big(e^{-\Gamma_+ t}
                       -e^{-\Gamma_- t}\Big)\ }\qquad\quad
\eqalign{
&M^{(1)}_{12}(t)={\widetilde D\over\Delta\Gamma}
                  \Big(e^{-\gamma_L t}-e^{-\gamma_S t}\Big)\cr
&M^{(1)}_{14}(t)={2\widetilde C^*\over\Delta\Gamma_-}\Big(
                 e^{-\gamma_S t}-e^{-\Gamma_+  t}\Big)\cr
&\cr
&M^{(1)}_{22}(t)=0 \phantom{\gamma\over\Delta\Gamma}\cr
&M^{(1)}_{24}(t)={2\widetilde C^*\over\Delta\Gamma_+}\Big(
                           e^{-\gamma_L t}-e^{-\Gamma_+  t}\Big)\cr
&\cr
&M^{(1)}_{32}(t)={2\widetilde C^*\over\Delta\Gamma_-}\Big(e^{-\gamma_L t}
-e^{-\Gamma_- t}\Big)\cr
&M^{(1)}_{34}(t)={i\widetilde B\over2\Delta m}\,
                   \Big(e^{-\Gamma_+ t}-e^{-\Gamma_- t}\Big)\cr
&\cr
&M^{(1)}_{42}(t)={2\widetilde C\over\Delta\Gamma_+}\Big(e^{-\gamma_L t}
                                 -e^{-\Gamma_+ t}\Big)\cr
&M^{(1)}_{44}(t)=0\ ,\phantom{i\widetilde B\over2\Delta m}}
\eqno(B.5)
$$
\line{}

\noindent
where $\Delta\Gamma_\pm=\Delta\Gamma\pm 2i\Delta m$.
In presenting the above expressions for the entries of 
$M^{(0)}$ and $M^{(1)}$, we have reconstructed the exponential dependences
out of first order correction terms which are linear 
in time.[21]

Although the time dependence in the solution of $(B.1)$§
is always exponential, this feature is lost in perturbation theory.
However, to a given order in the perturbative
expansion, one can always reconstruct the correct exponential behaviour
by a redefinition of the parameters $\gamma_S$, $\gamma_L$
and $\Delta m$, so that they coincide with the widths and mass difference of
the physical states $|B_S\rangle$ and $|B_L\rangle$. To first order, only
the widths get shifted: $\gamma_S\rightarrow \gamma_S +\widetilde  D$,
$\gamma_L\rightarrow \gamma_L +\widetilde  D$.
Accordingly, if we redefine $\Gamma$ to be the average
of the new $\gamma_S$ and $\gamma_L$, then the quantities
$\Gamma_\pm$ get changed as:
$\Gamma_\pm\rightarrow \Gamma_\pm +\widetilde A- \widetilde D$.
This explains the form of the exponential terms in $(B.4)$.

\vskip 2cm
%\vfill\eject

{\bf APPENDIX C}
\medskip

As explained in the text, the evolution in time of the two correlated
neutral $B$-mesons coming from the decay of the $\Upsilon(4S)$
resonance can be obtained using the single meson dynamics
$|\tilde\rho(0)\rangle\rightarrow|\tilde\rho(t)\rangle$
generated by the equation (3.16). 
It results convenient to regroup
the four components of the vector $|\tilde\rho(t)\rangle$
into a $2\times2$ matrix
$$
\tilde\rho(t)=\left[\matrix{\tilde\rho_1(t) & \tilde\rho_3(t)\cr
                            \tilde\rho_4(t) & \tilde\rho_2(t)}\right]\ ;
\eqno(C.1)
$$
it can be expressed in terms of the elementary operators
$\widetilde P_i$, $i=1,2,3,4$, introduced in (6.5):
$$
\tilde\rho(t)\equiv\sum_{i=1}^4 \widetilde P_i\ \tilde\rho_i(t)\ .
\eqno(C.2)
$$
In order to compute explicitly the form of the various observables
involving correlated mesons, we need to determine the time evolution
of the operators $\widetilde P_i$. Using the results of Appendix B,
on easily finds [adopting the same notation as in $(C.1)$]:
$$
\eqalign{
&\widetilde P_1(t)=\left[
\matrix{M^{(0)}_{11}(t) & M^{(1)}_{31}(t)\cr
        M^{(1)}_{41}(t) & M^{(1)}_{21}(t)}
\right]\cr
&\cr
&\widetilde P_3(t)=\left[
\matrix{M^{(1)}_{13}(t) &  M^{(0)}_{33}(t)\cr
        M^{(1)}_{43}(t) &  M^{(1)}_{23}(t)}
\right]
}\qquad\qquad
\eqalign{
&\widetilde P_2(t)=\left[
\matrix{M^{(1)}_{12}(t) & M^{(1)}_{32}(t)\cr
        M^{(1)}_{42}(t) & M^{(0)}_{22}(t)}
\right]\cr
&\cr
&\widetilde P_4(t)=\left[
\matrix{M^{(1)}_{14}(t) & M^{(1)}_{34}(t)\cr
        M^{(0)}_{44}(t) & M^{(1)}_{24}(t)}
\right]\ .}
\eqno(C.3)
$$
With these expressions, on can now calculate the elementary double probabilities
in (6.10)
$$
\eqalign{
{\cal P}_{ij}(t_1,t_2)
&=\hbox{Tr}\big\{\widetilde P_i\, \widetilde P_1(t_1)\big\}\
\hbox{Tr}\big\{\widetilde P_j\, \widetilde P_2(t_2)\big\}\ +\
\hbox{Tr}\big\{\widetilde P_i\, \widetilde P_2(t_1)\big\}\ 
\hbox{Tr}\big\{\widetilde P_j\, \widetilde P_1(t_2)\big\}\cr
&\qquad
-\hbox{Tr}\big\{\widetilde P_i\,  \widetilde P_3(t_1)\big\}\
\hbox{Tr}\big\{\widetilde P_j\, \widetilde P_4(t_2)\big\}\ -\
\hbox{Tr}\big\{\widetilde P_i\, \widetilde P_4(t_1)\big\}\
\hbox{Tr}\big\{\widetilde P_j\, \widetilde P_3(t_2)\big\}\ ,
}
\eqno(C.4)
$$
and, with a further time integration, the single-time probabilities
$$
{\mit\Pi}_{ij}(t)=\int_0^\infty dt'\  {\cal P}_{ij}(t'+t,t')\ .
\eqno(C.5)
$$
Introducing again the variable $\tau=\Gamma\, t$ and recalling the definitions
(5.3) and (5.6), one explicitly finds:

\vfill\eject

$$
\eqalign{
&{\mit\Pi}_{11}(\tau)={e^{-\tau}\over2\,\Gamma}\
{D\over(1+\delta)}\bigg[\cosh\delta\tau+{\sinh\delta\tau\over\delta}\bigg]\ ,\cr
&{\mit\Pi}_{12}(\tau)={e^{-\tau}\over2\,\Gamma}\ e^{-\delta\tau}\ ,\cr
&{\mit\Pi}_{13}(\tau)={e^{-\tau}\over2\,\Gamma}\
{C\over\delta+i\omega}\Big[e^{i\omega\tau} - r^*\, e^{-\delta\tau}\Big]\ ,\cr
&{\mit\Pi}_{14}(\tau)=\Big[{\mit\Pi}_{13}(\tau)\Big]^*\ ,\cr
&\cr
&{\mit\Pi}_{21}(\tau)={e^{-\tau}\over2\,\Gamma}\ e^{\delta\tau}\ ,\cr
&{\mit\Pi}_{22}(\tau)={e^{-\tau}\over2\,\Gamma}\
{D\over(1-\delta)}\bigg[\cosh\delta\tau+{\sinh\delta\tau\over\delta}\bigg]\ ,\cr
&{\mit\Pi}_{23}(\tau)={e^{-\tau}\over2\,\Gamma}\
{C\over\delta-i\omega}\bigg[e^{i\omega\tau}-{e^{\delta\tau}\over r}\bigg]\ ,\cr
&{\mit\Pi}_{24}(\tau)=\Big[{\mit\Pi}_{23}(\tau)\Big]^*\ ,\cr
&\cr
&{\mit\Pi}_{31}(\tau)={e^{-\tau}\over2\,\Gamma}\
{C\over\delta+i\omega}\Big[e^{\delta\tau} - r^*\, e^{-i\omega\tau} \Big]\ ,\cr
&{\mit\Pi}_{32}(\tau)={e^{-\tau}\over2\,\Gamma}\
{C\over\delta-i\omega}\bigg[e^{-\delta\tau}-{e^{-i\omega\tau}\over r}\bigg]\ ,\cr
&{\mit\Pi}_{33}(\tau)=-{e^{-\tau}\over2\,\Gamma}\
{B^*\over1+i\omega}\big[\omega\cos\omega\tau+\sin\omega\tau\big]\ ,\cr
&{\mit\Pi}_{34}(\tau)=-{e^{-\tau}\over2\,\Gamma}\
{e^{-(A-D+i\omega)\tau}\over 1+A-D}\ ,\cr
&\cr
&{\mit\Pi}_{41}(\tau)=\Big[{\mit\Pi}_{31}(\tau)\Big]^*\ ,\cr
&{\mit\Pi}_{42}(\tau)=\Big[{\mit\Pi}_{32}(\tau)\Big]^*\ ,\cr
&{\mit\Pi}_{43}(\tau)=\Big[{\mit\Pi}_{34}(\tau)\Big]^*\ ,\cr
&{\mit\Pi}_{44}(\tau)=\Big[{\mit\Pi}_{33}(\tau)\Big]^*\ ,\cr
}
\eqno(C.6)
$$
\line{}

\noindent
where the parameter $r$ represents the following ratio:
$$
r={2-\delta+i\omega \over 2+\delta-i\omega}\ .
\eqno(C.7)
$$
These expressions for the components of ${\mit\Pi}_{ij}(\tau)$
have been used to compute the double decay rates
${\mit\Gamma}(f_1,f_2;\tau)$ in (6.11).

%\vskip 2cm
\vfill\eject

\centerline{\bf REFERENCES}
\vskip 1cm

\item{1.} G.C. Branco, L. Lavoura and J.P. Silva, {\it CP Violation},
(Clarendon Press, Oxford, 1999)
\smallskip
\item{2.} I.I. Bigi and A.I. Silva, {\it CP Violation},
(Cambridge Univeristy Press, Cambridge, 2000)
\smallskip
\item{3.} P. Eerola, Nucl. Instr. and Meth. {\bf A446} (2000) 384
\smallskip
\item{4.} R. Alicki and K. Lendi, {\it Quantum Dynamical Semigroups and 
Applications}, Lect. Notes Phys. {\bf 286}, (Springer-Verlag, Berlin, 1987)
\smallskip
\item{5.} V. Gorini, A. Frigerio, M. Verri, A. Kossakowski and
E.C.G. Surdarshan, Rep. Math. Phys. {\bf 13} (1978) 149 
\smallskip
\item{6.} H. Spohn, Rev. Mod. Phys. {\bf 52} (1980) 569
\smallskip
\item{7.} A. Royer, Phys. Rev. Lett. {\bf 77} (1996) 3272
\smallskip
\item{8.} C.W. Gardiner and P. Zoller,
{\it Quantum Noise}, 2nd. ed. (Springer, Berlin, 2000)
\smallskip
\item{9.} W.H. Louisell, {\it Quantum Statistical Properties of Radiation},
(Wiley, New York, 1973)
\smallskip
\item{10.} M.O. Scully and M.S. Zubairy, 
{\it Quantum Optics} (Cambridge University Press, Cambridge, 1997)
\smallskip
\item{11.} L. Fonda, G.C. Ghirardi and A. Rimini, Rep. Prog. Phys.
{\bf 41} (1978) 587 
\smallskip
\item{12.} H. Nakazato, M. Namiki and S. Pascazio,
Int. J. Mod. Phys. {\bf B10} (1996) 247
\smallskip
\item{13.} F. Benatti and R. Floreanini, Phys. Lett. {\bf B428} (1998) 149
\smallskip
\item{14.} F. Benatti and R. Floreanini, Phys. Lett. {\bf B451} (1999) 422
\smallskip
\item{15.} F. Benatti and R. Floreanini, JHEP {\bf 02} (2000) 032
\smallskip
\item{16.} F. Benatti and R. Floreanini, Phys. Rev. D {\bf 62} (2000) 125009
\smallskip
\item{17.} A.D. Dolgov, Sov. J. Nucl. Phys. {\bf 33} (1981) 700
\smallskip
\item{18.} G. Sigl and G. Raffelt, Nucl. Phys. {\bf B406} (1993) 423
\smallskip
\item{19.} J. Ellis, J.L. Lopez, N.E. Mavromatos 
and D.V. Nanopoulos, Phys. Rev. D {\bf 53} (1996) 3846
\smallskip
\item{20.} P. Huet and M.E. Peskin, Nucl. Phys. {\bf B434} (1995) 3
\smallskip
\item{21.} F. Benatti and R. Floreanini, Nucl. Phys. {\bf B488} (1997) 335
\smallskip
\item{22.} F. Benatti and R. Floreanini, Phys. Lett. {\bf B401} (1997) 337
\smallskip
\item{23.} F. Benatti and R. Floreanini, Nucl. Phys. {\bf B511} (1998) 550
\smallskip
\item{24.} F. Benatti and R. Floreanini, $CPT$, dissipation, and all that,
in {\it Physics and Detectors for Daphne}, ``Frascati Physics Series'', vol. XVI,
1999, p. 307, {\tt hep-ph/9912426}
\smallskip
\item{25.} F. Benatti and R. Floreanini, Phys. Lett. {\bf B465} (1999) 260
\smallskip
\item{26.} M.S. Marinov, JETP Lett. {\bf 15} (1972) 479; Sov. J. Nucl. Phys.
{\bf 19} (1974) 173; Nucl. Phys. {\bf B253} (1985) 609
\smallskip
\item{27.} S. Hawking, Comm. Math. Phys. {\bf 87} (1983) 395; Phys. Rev. D
{\bf 37} (1988) 904; Phys. Rev. D {\bf 53} (1996) 3099;
S. Hawking and C. Hunter, Phys. Rev. D {\bf 59} (1999) 044025
\smallskip
\item{28.} J. Ellis, J.S. Hagelin, D.V. Nanopoulos and M. Srednicki,
Nucl. Phys. {\bf B241} (1984) 381; 
\smallskip
\item{29.} S. Coleman, Nucl. Phys. {\bf B307} (1988) 867
\smallskip
\item{30.} S.B. Giddings and A. Strominger, Nucl. Phys. {\bf B307} (1988) 854
\smallskip
\item{31.} M. Srednicki, Nucl. Phys. {\bf B410} (1993) 143
\smallskip
\item{32.} W.G. Unruh and R.M. Wald, Phys. Rev. D {\bf 52} (1995) 2176
\smallskip
\item{33.} L.J. Garay, Phys. Rev. Lett. {\bf 80} (1998) 2508;
Phys. Rev. D {\bf 58} (1998) 124015
\smallskip
\item{34.} J. Ellis, N.E. Mavromatos and D.V. Nanopoulos, Phys. Lett.
{\bf B293} (1992) 37; Int. J. Mod. Phys. {\bf A11} (1996) 1489
\smallskip
\item{35.} F. Benatti and R. Floreanini, Ann. of Phys. {\bf 273} (1999) 58
\smallskip
\item{36.} V.A. Kosteleck\'y and R. Van Kooten, Phys. Rev. D {\bf 54}
(1996) 5585
\smallskip
\item{37.} P. Colangelo and G. Corcella, Eur. Phys. J. C {\bf 1} 
(1998) 515
\smallskip
\item{38.} S. Yang and G. Isidori, Test of $CPT$ invariance in semileptonic
$B$ decays, BaBar \hbox{Note \#438}, 1998
\smallskip
\item{39.} R.A. Bertlmann and W. Grimus, Phys. Rev. D {\bf 58} (1998) 034014
\smallskip
\item{40.} A. Mohapatra, M. Satpathy, K. Abe and Y. Sakai,
Phys. Rev. D {\bf 58} (1998) 036003
\smallskip
\item{41.} L. Lavoura and J.P. Silva, Phys. Rev. D {\bf 60} (1999) 056003
\smallskip
\item{42.} F. Benatti and R. Floreanini,
Mod. Phys. Lett. {\bf A12} (1997) 1465; 
Banach Center Publications, {\bf 43} (1998) 71; 
Phys. Lett. {\bf B468} (1999) 287; On the weak-coupling limit and complete
positivity, Chaos Sol. Frac., to appear
\smallskip
\item{43.} F. Benatti and R. Floreanini, J. Phys. {\bf A33} (2000) 8139
\smallskip
\item{44.} N.W. Tanner and R.H. Dalitz, Ann. of Phys. {\bf 171} (1986) 463
\smallskip
\item{45.} M.C. Banuls and J. Bernabeu, Phys. Lett. {\bf B423} (1998) 151;
{\bf B464} (1999) 117; Nucl Phys. {\bf B590} (2000) 19
\smallskip
\item{46.}  L. Lavoura, Phys. Lett. {\bf B442} (1998) 390
\smallskip
\item{47.} The OPAL Collaboration, Zeit. fur Physik {\bf C76} (1997) 401
\smallskip
\item{48.} The OPAL Collaboration, Eur. Phys. J. {\bf C12} (2000) 609 
\smallskip
\item{49.} P. Ball {\it et al.}, $B$ decays at the LHC,
CERN-TH-2000-101, {\tt hep-ph/0003238}
\smallskip
\item{50.} The BTeV proposal, 2000, 
\hbox{\tt http://www-btev.fnal.gov}
\smallskip
\item{51.} The BaBar Collaboration, {\it The BaBar Physics Book}, 
SLAC-R-504, 1998
\smallskip
\item{52.} G. De Domenico and Ch. Y\`eche, Dilepton analysis in BaBar
experiment: measurement of the mixing parameter $\Delta m_B$ and study
of the $T$ ($CP$) violation purely in mixing, BaBar Note \#409, 1998
\smallskip
\item{53.} C. Leonidopoulos, $B^0$-$\overline{B^0}$ mixing and $CPT$
violation with the Belle experiment, Ph.D. thesis, Princeton University,
2000
\smallskip
\item{54.} The CLEO Collaboration, Phys. Lett. {\bf B490} (2000) 37
\smallskip
\item{55.} Particle Data Group, Eur. Phys. J. C {\bf 15} (2000) 1
\smallskip
\item{56.} S. Adler, Phys. Rev. D {\bf 62} (2000) 117901
\smallskip
\item{57.} J. Ellis, N.E. Mavromatos and D.V. Nanopoulos,
Phys. Rev. D {\bf 63} (2001) 024024

\bye